\documentclass[preprint,preprintnumbers,amsmath,amssymb]{revtex4}
\usepackage{graphicx}% Include figure files
\usepackage{dcolumn}% Align table columns on decimal point
\usepackage{bm}% bold math
\usepackage{epsfig}

\def\integral{\int_{-\infty}^{+\infty}}

\begin{document}
\DeclareGraphicsRule{.jpg}{eps}{.jpg.bb}{`jpeg2ps #1}
\hfill {\bf December 2003}

\title{Wigner quasi-probability distribution for the infinite
square well: energy eigenstates and time-dependent wave packets}

\author{M. Belloni} \email{mabelloni@davidson.edu}
\affiliation{%
Physics Department \\
Davidson College \\
Davidson, NC 28035 USA \\
}

\author{M. A. Doncheski} \email{mad10@psu.edu}
\affiliation{%
Department of Physics\\
The Pennsylvania State University \\
Mont Alto, PA 17237 USA \\
}

\author{R. W. Robinett} \email{rick@phys.psu.edu}
\affiliation{%
Department of Physics\\
The Pennsylvania State University\\
University Park, PA 16802 USA \\
}

\date{\today}

\begin{abstract}
We calculate the Wigner quasi-probability distribution for position
and momentum, $P_{W}^{(n)}(x,p)$, for the energy eigenstates of the standard
infinite well potential, using both $x$- and $p$-space stationary-state
solutions, as well as visualizing the results. We then evaluate the
time-dependent Wigner distribution, $P_{W}(x,p;t)$, for Gaussian wave packet
solutions of this system, illustrating both the short-term semi-classical
time dependence, as well as longer-term revival and fractional revival
behavior and the structure during the collapsed state.  This tool provides
an excellent way of demonstrating the patterns of highly correlated
Schr\"{o}dinger-cat-like `mini-packets' which appear at fractional multiples
of the exact revival time.
\end{abstract}

\maketitle

\section{Introduction}
\label{sec:introduction}

The solution and visualization of problems in one-dimensional
quantum mechanics focuses most often on calculations of the
position-space wavefunction, $\psi(x,t)$. More occasionally, problems
may be solved using, or transformed into, the momentum-space counterpart,
$\phi(p,t)$, using the standard Fourier transforms,
\begin{equation}
\psi(x,t)  =   \frac{1}{\sqrt{2\pi \hbar}}
\int_{-\infty}^{+\infty} \,\phi(p,t)\, e^{+ipx/\hbar}\,dp
\label{fourier_1} \\
\end{equation}
and
\begin{equation}
\phi(p,t)  =  \frac{1}{\sqrt{2\pi \hbar}}
\int_{-\infty}^{+\infty} \,\psi(x,t)\, e^{-ipx/\hbar}\,dx
\label{fourier_2}
\,.
\end{equation}
\\
Connections between the classical and quantum descriptions of
model systems can then be made in a variety of ways. For example,
the quantum mechanical expectation values, $\langle x \rangle_t$
and $\langle p \rangle_t$ can be compared to their classical
analogs, $x(t)$ and $p(t) = mv(t)$, via Ehrenfest's principle,
e.g., $\langle p \rangle_t = m d\langle x\rangle_t/dt$. Quantum
mechanical probability densities, $P_{QM}^{(n)}(x) =
|\psi_n(x)|^2$ and  $P_{QM}^{(n)}(p) = |\phi_n(p)|^2$, can be
related to classical probability distributions, calculated using
simple ``{\it how long does a particle spend in a given $dx$ or
$dp$ interval?}'' arguments, or justified in more detail through
the WKB approximation.

The visualization of solutions to problems in classical mechanics
through a phase-space description, i.e., parametric plots of
$p(t)$ versus $x(t)$, is often helpful.  It is natural to wonder
if a quantum mechanical analog of a phase-space probability
distribution, a joint $P(x,p)$ probability density,  is a useful
construct, despite the obvious problems raised by the Heisenberg
uncertainty principle and its connection between $x$ and $p$.

Wigner \cite{wigner} was one of the first to address this issue and
introduced a quasi- or pseudo-probability distribution corresponding
to a general quantum state, $\psi(x,t)$, defined by
\begin{equation}
P_{W}(x,p;t)
\equiv
\frac{1}{\pi \hbar}
\int_{-\infty}^{+\infty}
\psi^{*}(x+y,t)\,\psi(x-y,t)\,e^{2ipy/\hbar}\,dy
\, .
\end{equation}
The properties of the Wigner distribution have been discussed in a
number of accessible reviews \cite{tatarskii} - \cite{weyl_review},
analyzed for theoretical consistency \cite{1981_wigner} - \cite{wheeler},
used for a number of physical applications
\cite{first_dahl} - \cite{schleich_research},
discussed in the pedagogical literature \cite{snygg} - \cite{campos},
and have even made the occasional brief appearance in quantum
mechanics textbooks \cite{marek}, \cite{ballentine}.
Similar topics are often discussed in the context of both the signal
analysis of time-varying spectra \cite{cohen} (where the two complementary
variables are $t$ and $\omega$) and quantum
optics \cite{scully}, \cite{schleich}.

Of all of the familiar model systems often used as tractable
pedagogical examples for the application of quantum mechanical
techniques, many of the standard potentials have been analyzed in
detail using the Wigner distribution, including the harmonic
oscillator, linear potential, Morse potential, hydrogen atom
(Coulomb problem), and others. The most familiar example of all,
however, the infinite square well (ISW) problem, has received less
attention \cite{lee}, \cite{weyl_review}, \cite{casas}. The Wigner
function has recently been used to help explain the interesting
patterns of probability density, $P(x,t) = |\psi(x,t)|^2$ versus
$(x,t)$, when plotted over long time periods (namely, over the
interval $0<t<T_{rev}$ where $T_{rev}$ is the revival time, to be
discussed below). Such patterns have come to be known as quantum
carpets \cite{quantum_kinzel} - \cite{other_carpets}.   The
authors of Ref.~\cite{marzoli} have produced video sequences of
the free evolution of an array of Wigner functions constructed
from an antisymmetric wavefunction of a wave packet and its mirror
wave packet in order to understand the `weaving' of quantum
carpets.  While such graphical representations of the long-term
time dependence of this familiar problem are very interesting, the
necessary background material which would allow students (or
instructors) to approach such problems themselves has not been
stressed.

The purpose of this paper, therefore, is to thoroughly review the derivation
of the Wigner distribution for this very accessible problem, not only for
individual energy eigenstates, but for time-dependent wave packet solutions.
This last case is important as the Wigner function provides a very useful tool
for the visualization of the non-trivial long-term time evolution of
initially localized wave packets, including revival and
fractional revival behaviors.
(For a pedagogical review of this topic, see Ref.~\cite{bluhm_ajp},
and for a recent survey of the research literature,
see Ref.~\cite{robinett_physics_report}.)

In the next section, we provide a short, self-contained review of many
of the general properties of the Wigner quasi-probability distribution,
followed in Section~\ref{sec:infinite_well_results} by a collection of
useful results for the ISW energy eigenfunctions, in both position and
momentum space, helpful for the evaluation of the Wigner function for
the ISW.  We briefly review the classical phase-space description of the
infinite square well in Sec.~\ref{sec:phase_space_isw}, and
in Section~\ref{sec:wigner_isw_1} we exhibit the Wigner
distribution, $P_{W}^{(n)}(x,p)$, for the ISW eigenstates, illustrating
their derivations in both $x$- and $p$-space, and comparing them to
their classical counterparts. In  Section~\ref{sec:wigner_isw_2},  we
extend these results to general wave packet solutions of the infinite square
well, focusing on the structure of the Wigner function for time-dependent
Gaussian wave packet solutions of the ISW, at a number of full and fractional
revivals, as well as during the so-called collapsed phase, illustrating
the power of this method for the visualization of time-dependent
quantum phenomena. Throughout, we focus on providing explicit analytic
results which can be used for further numerical or visual
investigations.

\section{Review of the Wigner quasi-probability distribution}
\label{sec:wigner_distribution}

The definition of the Wigner function,
\begin{equation}
P_{W}(x,p;t)
\equiv
\frac{1}{\pi \hbar}
\int_{-\infty}^{+\infty}
\psi^{*}(x+y,t)\,\psi(x-y,t)\,e^{2ipy/\hbar}\,dy
\, ,
\label{wigner_definition_x}
\end{equation}
using position-space wavefunctions, can be easily rewritten in
momentum space, using Eqn.~(\ref{fourier_2}), in the equivalent form
\begin{equation}
P_{W}(x,p;t) =
\frac{1}{\pi \hbar}
\int_{-\infty}^{+\infty}
\phi^*(p+q,t)\, \phi(p-q,t)\, e^{-2iqx/\hbar}\,dq
\,.
\label{wigner_definition_p}
\end{equation}
The Wigner distribution is easily shown to be real,
\begin{eqnarray}
\left[P_{W}(x,p;t)\right]^{*}
& = &
\frac{1}{\pi \hbar}
\int_{-\infty}^{+\infty}
\psi(x+y,t)\, \psi^{*}(x-y,t)\, e^{-2ipy/\hbar}\,dy
\nonumber \\
& = &
\frac{1}{\pi \hbar}
\int_{-\infty}^{+\infty}
\psi^{*}(x+\overline{y},t)\, \psi(x-\overline{y},t)
\, e^{+2ip\overline{y}/\hbar}\,d\overline{y} \\
& = & P_{W}(x,p;t) \nonumber
\end{eqnarray}
by using a simple change of variables ($\overline{y} = -y$).  This is,
of course, one of the desired properties of a probability distribution.

Integration of $P_{W}(x,p;t)$ over one variable or the other is seen to
give the correct marginal probability distributions for $x$ and $p$
separately, since
\begin{eqnarray}
\int_{-\infty}^{+\infty}
P_{W}(x,p;t)\, dp & = & |\psi(x,t)|^2 = P_{QM}(x,t)
\label{integrate_p} \\
\int_{-\infty}^{+\infty}
P_{W}(x,p;t)\, dx & = & |\phi(p,t)|^2 = P_{QM}(p,t)
\label{integrate_x}
\end{eqnarray}
where one uses the definition of the Dirac $\delta$ function
\begin{equation}
\delta(z) = \frac{1}{2\pi} \int_{-\infty}^{+\infty}\,e^{ikz}\,dk
\end{equation}
in Eqns.~(\ref{wigner_definition_x}) or (\ref{wigner_definition_p}).
This property of $P_{W}(x,p)$ is clearly another necessary condition
for a joint probability density.

However, one can also easily show that the
Wigner distributions for two distinct quantum states, $\psi(x,t)$ and
$\chi(x,t)$,
\begin{eqnarray}
P_{W}^{(\psi)}(x,p;t) & = & \frac{1}{\pi \hbar} \int_{-\infty}^{+\infty}
\psi^{*}(x+y,t)\,\psi(x-y,t)\,e^{2ipy/\hbar}\,dy \\
P_{W}^{(\chi)}(x,p;t) & = & \frac{1}{\pi \hbar} \int_{-\infty}^{+\infty}
\chi^{*}(x+z,t)\,\chi(x-z,t)\,e^{2ipz/\hbar}\,dz
\end{eqnarray}
satisfy the relation
\begin{equation}
\int_{-\infty}^{+\infty}\,dx
\,
\int_{-\infty}^{+\infty}\, dp
\,
P_{W}^{(\psi)}(x,p;t)
\,
P_{W}^{(\chi)}(x,p;t)
= \frac{2}{\pi \hbar}
|\langle \psi | \chi \rangle|^2
\,.
\label{orthogonal_integral}
\end{equation}
So, for example, if $\psi$ and $\chi$ are orthogonal states so that
$\langle \psi | \chi \rangle =0$, it cannot be true that the corresponding
Wigner distributions can be everywhere non-negative, as there must be
cancellations in the integral in Eqn.~(\ref{orthogonal_integral}). The
fact that $P_{W}(x,p;t)$ can be negative is easily confirmed by direct
calculation, for example, with simple cases such as the harmonic oscillator
(for any $n \geq 1$ state). This feature is, after all, hardly surprising
because of the non-commutativity of $x$ and $p$ encoded in the uncertainty
principle.  Despite this apparent drawback \cite{drawback}, the Wigner
function is still useful for the visualization of the correlated position- and
momentum-space behavior of quantum eigenstates and wave packets.

A useful benchmark example of the Wigner function is for a
Gaussian free-particle wave packet, where the time-dependent
momentum- and position-space wavefunctions (for arbitrary initial
$x_0$ and $p_0$) can be written in the forms
\begin{eqnarray}
\phi_{(G)}(p,t)
& = &
\sqrt{\frac{\alpha}{\sqrt{\pi}}}
\, e^{-\alpha^2(p-p_0)^2/2}
\, e^{-ipx_0/\hbar}
\, e^{-ip^2t/2m\hbar}
\label{p_gaussian_t} \\
\psi_{(G)}(x,t) & = &  \frac{1}{\sqrt{\sqrt{\pi} \alpha \hbar (1+it/t_0)}}
\,
e^{ip_0(x-x_0)/\hbar}
\, e^{-ip_0^2t/2m\hbar}
\,
e^{-(x-x_0-p_{0}t/m)^2/2(\alpha \hbar)^2(1+it/t_0)}
\label{x_gaussian_t}
\end{eqnarray}
which are related by Eqns.~(\ref{fourier_1}) and (\ref{fourier_2}).
These solutions are, of course, well-known to be characterized by
\begin{eqnarray}
\langle p \rangle_t = p_0,
& \quad  \quad&
\Delta p_t = \frac{1}{\alpha \sqrt{2}} \\
\langle x \rangle_t = (p_0/m)t + x_0,
& \quad  \quad &
\Delta x_t = (\hbar \alpha/\sqrt{2}) \sqrt{1 + (t/t_0)^2}
\end{eqnarray}
where $t_0 \equiv m\hbar \alpha^2$ is the spreading time.
The corresponding Wigner function is easily obtained \cite{kim_noz},
using standard Gaussian integrals, and is given by
\begin{equation}
P_{W}^{(G)}(x,p;t) = \frac{1}{\pi \hbar}
\,
e^{-\alpha^2 (p-p_0)^2}
\,
e^{-(x-x_0-pt/m)^2/\beta^2}
\label{free_particle_gaussian_wigner}
\end{equation}
where $\beta \equiv \hbar \alpha$. In this case, the ultra-smooth
Gaussian solution does give a positive-definite $P_{W}(x,p;t)$ and it
is known \cite{hudson} that such solutions are the only forms which
give rise to non-negative Wigner functions.

Another result which will prove useful in what follows is the
expression for the Wigner function for the case of a linear combination
of two 1-D Gaussians, characterized by different values of $x_0$ and $p_0$.
For example, if we assume that at some instant of time we have
\begin{eqnarray}
\psi^{(A,B)}(x) & = & \gamma \psi_{(G)}(x;x_A,p_A) +
\delta  \psi_{(G)}(x;x_B,p_B)
 \\
&  = &
\gamma
\left[
\frac{1}{\sqrt{\beta\sqrt{\pi}}}
e^{-(x-x_A)^2/2\beta^2}e^{ip_A(x-x_A)/\hbar}
\right]
+
\delta
\left[
\frac{1}{\sqrt{\beta\sqrt{\pi}}}
e^{-(x-x_B)^2/2\beta^2}e^{ip_B(x-x_B)/\hbar}
\right] \nonumber
\, ,
\end{eqnarray}
the corresponding Wigner function is given by
\begin{eqnarray}
P_{W}^{(A,B)}
& = &  \frac{1}{\pi \hbar}
\left[
|\gamma|^2
e^{-\alpha^2(p-p_A)^2}e^{-(x-x_A)^2/\beta^2}
+
|\delta|^2
e^{-\alpha^2(p-p_B)^2}e^{-(x-x_B)^2/\beta^2} \right.
\label{two_gaussian_wigner} \\
& &
\qquad
\left.
+
2e^{-\alpha^2(p-\overline{p})^2}e^{-(x-\overline{x})^2/\beta^2}
Re\left\{\gamma \delta^* e^{i(x_A p_B - x_B p_A)/\hbar}
\,
e^{-i(x_A-x_B)(p-\overline{p})/\hbar}
\,
e^{i(p_A-p_B)x/\hbar}
\right\}
\right]
\nonumber
\end{eqnarray}
where
\begin{equation}
\overline{x} \equiv
\frac{x_A+x_B}{2}
\qquad
\mbox{and}
\qquad
\overline{p}
\equiv
\frac{p_A+p_B}{2}
\, .
\label{average_values}
\end{equation}
In this case, the Wigner function is characterized by two smooth `lumps'
in phase space, corresponding to the values of $(x_A,p_A)$ and $(x_B,p_B)$
of the individual Gaussians, but also by an oscillatory term, centered at a
point in phase space defined by the average of these values; the
oscillations appear in the $x$ variable (if $p_A \neq p_B$),
the $p$ variable (if $x_A \neq x_B$) or both. (A similar expression
for the case where $p_A = p_B = 0$ is given in Ref.~\cite{ballentine};
see also  Refs.~\cite{cohen} and \cite{schleich} for related examples
involving time-frequency analysis and coherent states, respectively.)

For a single stationary state or energy-eigenfunction solution, corresponding
to a quantized bound state, given by
\begin{equation}
\psi_{n}(x,t) = u_{n}(x) e^{-iE_nt/\hbar}
\end{equation}
where $u_n(x)$ can be considered as a purely {\it real} function,
the Wigner distribution is time independent
\begin{equation}
P_{W}^{(n)}(x,p;t) =
\frac{1}{\pi \hbar}
\int_{-\infty}^{+\infty} u_n(x+y)\, u_n(x-y)\, e^{2ipy/\hbar}\,dy
= P_{W}^{(n)}(x,p)
\, .
\end{equation}
For such solutions, another change of variables argument suffices
to show that
\begin{equation}
P_{W}^{(n)}(x,-p) = P_{W}^{(n)}(x,+p)
\end{equation}
which corresponds to the classical result that a particle undergoing
bound state motion will spend equal amounts of time with  $+p$
(motion to the right) as in the $-p$ direction (going to the left).
A more familiar version of this is often discussed in textbooks,
where for real eigenstates one knows that
\begin{eqnarray}
\phi_{n}(p)
& = &
\frac{1}{\sqrt{2\pi \hbar}}
\int_{-\infty}^{+\infty} u_n(x)\, e^{-ipx/\hbar}\,dx \\
& = &
\frac{1}{\sqrt{2\pi \hbar}}
\int_{-\infty}^{+\infty} u_n(x)\, [\cos(px/\hbar) -i \sin(px/\hbar)]\,dx \\
& \equiv & A(p) - iB(p)
\end{eqnarray}
where $A(p),B(p)$ are then purely real functions. This immediately implies
that
\begin{equation}
|\phi_{n}(-p)|^2 = |A(p)+iB(p)|^2
= [A(p)]^2 + [B(p)]^2 =
|A(p)- iB(p)|^2 = |\phi_{n}(+p)|^2
\, .
\end{equation}

The general expression for a wave packet solution constructed from such
energy eigenstates is
\begin{equation}
\psi(x,t) = \sum_{n=1}^{\infty} a_n \,u_n(x) \, e^{-iE_nt/\hbar}
\end{equation}
where the expansion coefficients satisfy $\sum_n |a_n|^2 =1$. The
time-dependent Wigner distribution for this case is then
\begin{eqnarray}
P_{W}^{(\psi)}(x,p;t)
& \equiv &
\frac{1}{\pi \hbar}
\int_{-\infty}^{+\infty}
\psi^{*}(x+y,t)\,\psi(x-y,t)\,e^{2ipy/\hbar}\,dy
\nonumber \\
& = &
\sum_{m=1}^{\infty}
\sum_{n=1}^{\infty}
[a_m]^*a_n \, e^{i(E_m-E_n)t/\hbar}\,
\left[
\frac{1}{\pi \hbar}
\int_{-\infty}^{+\infty}
u_m(x+y)\,u_n(x-y)\,e^{2ipy/\hbar}\,dy \right] \nonumber \\
& \equiv &
\sum_{m=1}^{\infty}
\sum_{n=1}^{\infty}
[a_m]^*a_n \, e^{i(E_m-E_n)t/\hbar}\,
P_{W}^{(m,n)}(x,p) \label{time_dependent_wigner}
\end{eqnarray}
where, in general, we must calculate both diagonal ($m=n$) and
off-diagonal ($m\neq n$) terms of the form
\begin{equation}
P_{W}^{(m,n)}(x,p)
= \frac{1}{\pi \hbar}
\int_{-\infty}^{+\infty}
u_m(x+y)\,u_n(x-y)\,e^{2ipy/\hbar}\,dy
\,.
\label{off_diagonal}
\end{equation}
While $P_{W}^{(\psi)}(x,p;t)$ is real, the off-diagonal Wigner
terms are not necessarily so, but do satisfy
\begin{equation}
\left[P_{W}^{(m,n)}(x,p)\right]^*
=
P_{W}^{(n,m)}(x,p)
\, .
\end{equation}

\section{Infinite square well results in position and momentum space}
\label{sec:infinite_well_results}

Because the Wigner function can be evaluated using either position- or
momentum-space wavefunctions, we review the properties of, and
interconnections between, these solutions for the infinite square well (ISW).
The standard problem of a particle of mass $m$ confined by the potential
\begin{equation}
      V(x) = \left\{ \begin{array}{ll}
               0 & \mbox{for $0<x<L$} \\
               \infty & \mbox{otherwise}
                                \end{array}
\right.
\label{infinite_well_potential}
\end{equation}
has energy eigenvalues and position-space eigenfunctions
(which are non-zero only within the well) which  are given by
\begin{equation}
E_n = \frac{\hbar ^2 \pi^2 n^2}{2mL^2}
= \frac{p_n^2}{2m} \qquad \mbox{($p_n \equiv  n\pi \hbar/L$)}
\qquad
\mbox{with}
\qquad
u_n(x) = \sqrt{\frac{2}{L}} \sin\left(\frac{n\pi x}{L}\right)
\, .
\label{1d_eigenfunctions}
\end{equation}
The position-space eigenfunctions have a generalized parity property about
the midpoint of the well since
\begin{equation}
u_{n}(L-x) =
\sqrt{\frac{2}{L}} \sin\left(\frac{n\pi (L-x)}{L}\right)
=
-\sqrt{\frac{2}{L}} \cos(n\pi) \sin\left(\frac{n\pi x}{L}\right)
=
(-1)^{n+1} \, u_{n}(x)
\,.
\label{general_parity}
\end{equation}
As with any such system, the eigenfunctions can be made orthonormal,
which in this case can be readily checked explicitly by direct
calculation of
\begin{eqnarray}
\langle u_m | u_n \rangle
& =  &
\left(\frac{2}{L}\right)
\int_{0}^{L} \sin\left(\frac{m\pi x}{L}\right)\,
\sin\left(\frac{n\pi x}{L}\right)\, dx \nonumber \\
& = &  \left\{
\frac{\sin[(m-n)\pi]}{(m-n)\pi}
-
\frac{\sin[(m+n)\pi]}{(m+n)\pi}
\right\}
=
\delta_{m,n}
\, .
\label{orthonormality}
\end{eqnarray}
The momentum-space eigenfunctions  are given by Eqn.~(\ref{fourier_2})
as
\begin{eqnarray}
\phi_{n}(p) & = &
\frac{1}{\sqrt{2\pi \hbar}}
\int_{0}^{L} \,u_{n}(x)\, e^{-ipx/\hbar}\,dx \nonumber \\
& = &
(-i) \sqrt{\frac{L}{\pi \hbar}}
e^{-ipL/2\hbar}
\left[
e^{+in\pi/2}
\,
\frac{\sin[(pL/\hbar - n\pi)/2]}{(pL/\hbar - n\pi)}
-
e^{-in\pi/2}
\,
\frac{\sin[(pL/\hbar + n\pi)/2]}{(pL/\hbar + n\pi)}
\right] \nonumber
\end{eqnarray}
and the resulting probability density is given by
\begin{eqnarray}
|\phi_{n}(p)|^2
& = &
\left(\frac{L}{\hbar \pi}\right)
\left[
\frac{\sin^2[(pL/\hbar - n\pi)/2]}{(pL/\hbar - n\pi)^2}
+
\frac{\sin^2[(pL/\hbar + n\pi)/2]}{(pL/\hbar + n\pi)^2} \right.
\nonumber \\
& & \qquad \qquad \left. -
2\cos(n\pi)
\frac{\sin[(pL/\hbar - n\pi)/2]\sin[(pL/\hbar
+  n\pi)/2]}{(pL/\hbar - n\pi)(pL/\hbar + n\pi)}
\right]
\label{momentum_space_probability}
\end{eqnarray}
which will be  useful for visualization purposes.

While these forms demonstrate more explicitly the strong peaking
of the probability amplitude near the expected values of $p =
\pm p_n = \pm n \pi \hbar/L$, they do not make clear the finite extent
(limited to the range $[0,L]$) of the position-space amplitude. In order to
exemplify this dependence, we can evaluate the inverse Fourier transform,
via Eqn.~(\ref{fourier_1}), to explicitly obtain $u_n(x)$.  Because
such calculations will be useful in the evaluation of the Wigner
distribution using momentum-space wavefunctions, we examine this
seldom-discussed analysis \cite{lee} in some detail.
We require
\begin{eqnarray}
u_{n}(x)  & = &  \frac{1}{\sqrt{2\pi \hbar}}
\int_{-\infty}^{+\infty} \,\phi_{n}(p)\, e^{ipx/\hbar}\,dp \nonumber \\
& = & (-i) \sqrt{\frac{L}{2\pi^2 \hbar^2}}
\left\{
e^{+in\pi/2}
\int_{-\infty}^{+\infty} e^{ip(x-L/2)/\hbar}\,
\,
\frac{\sin[(pL/\hbar - n\pi)/2]}{(pL/\hbar - n\pi)} \,dp \right. \\
& &
\qquad \qquad \qquad \qquad
\left.
-
e^{-in\pi/2}
\int_{-\infty}^{+\infty} e^{ip(x-L/2)/\hbar}\,
\,
\frac{\sin[(pL/\hbar + n\pi)/2]}{(pL/\hbar + n\pi)}\,dp
\right\} \nonumber \\
& \equiv &
(-i) \sqrt{\frac{L}{2\pi^2 \hbar^2}}
\left\{ e^{+in\pi/2} I_A - e^{-in\pi/2} I_B\right\}
\nonumber
\,.
\end{eqnarray}
Each integral, $I_A,I_B$, can be done in turn using a change of variables,
giving, for example,
\begin{eqnarray}
I_A & = & \int_{-\infty}^{+\infty} e^{ip(x-L/2)/\hbar}\,
\, \frac{\sin[(pL/\hbar - n\pi)/2]}{(pL/\hbar - n\pi)} \, dp \nonumber \\
& = &
\left(\frac{\hbar}{L}\right) \,e^{+in\pi x/L}\,e^{-in\pi/2}
\,
\int_{-\infty}^{+\infty} \frac{\sin(q)}{q}\, e^{imq}\,dq \\
& = &
\left(\frac{\hbar}{L}\right)\, e^{+in\pi x/L}\,e^{-in\pi/2}
\,
\int_{-\infty}^{+\infty} \frac{\sin(q)}{q}\, [\cos(mq) +
i\sin(mq)] \,dq \nonumber
\label{intermediate_integral}
\end{eqnarray}
where we have used
\begin{equation}
q \equiv \frac{1}{2}\left(\frac{pL}{\hbar} - n\pi \right)
\qquad
\mbox{and}
\qquad
m \equiv \frac{(2x-L)}{L}
\, .
\label{first_change_of_variables}
\end{equation}
The piece of $I_A$ which includes the $\sin(mq)$ term vanishes
for symmetry reasons (odd integrand over a symmetric interval), while
the component with $\cos(mq)$ can be done using standard handbook
\cite{math_handbook} results, which we review in
Appendix~\ref{appendix:trig_integrals}.
The result is
\begin{equation}
\int_{-\infty}^{+\infty} \frac{\sin(q)}{q}\, \cos(mq)\,dq
          = \left\{ \begin{array}{ll}
             \pi  & \mbox{for $m^2<1  $} \\
              \pi/2 & \mbox{for $m^2=1  $} \\
              0 & \mbox{for $m^2>1$}
\end{array} \right. 
\label{important_integral}
\end{equation}
The restriction on $m$ corresponds to
\begin{equation}
m^2 = \left(\frac{2x-L}{L}\right)^2 < 1
\qquad
\longrightarrow
\qquad
-1 < \frac{2x-L}{L} < +1
\qquad
\longrightarrow
\qquad
0<x<L
\label{restrictions}
\end{equation}
which explicitly demonstrates the finite extent of the position-space
wavefunction which is non-vanishing only in the range $[0,L]$, as
expected. Performing the second integral ($I_B$) in the same way, and
combining factors, we find that the non-vanishing
position-space wavefunction is given by
\begin{eqnarray}
u_n(x)
& = &
(-i) \sqrt{\frac{L}{2\pi^2 \hbar^2}} \left(\frac{\hbar \pi}{L}
\right)
\left\{
e^{+in\pi/2} e^{+in\pi x/L}\, e^{-i n\pi/2}
-
e^{-in\pi/2} e^{-in\pi x/L}\, e^{+i n\pi/2}
\right\} \nonumber \\
& = &
\sqrt{\frac{2}{L}} \sin\left(\frac{n\pi x}{L}\right)
\qquad
\qquad
\qquad
\mbox{for $0<x<L$}
\end{eqnarray}
again, as expected. For future notational convenience, we can write this
in the form
\begin{equation}
u_{n}(x) = \sqrt{\frac{2}{L}} \sin\left(\frac{n\pi x}{L}\right)
\, {\cal R}(x;0,L)
\label{not_heaviside}
\end{equation}
where
\begin{equation}
{\cal R}(x;a,b) = \left\{ \begin{array}{ll}
0   &  \mbox{for $x<a,x>b$} \\
1/2 &  \mbox{for $x=a, x=b$} \\
1   &  \mbox{for $a<x<b$}
\end{array} \right. 
\, . \label{restricted_x_function}
\end{equation}
(We note that the results from Eqns.~(\ref{important_integral}), 
(\ref{not_heaviside}), and (\ref{restricted_x_function}) can also 
be expressed in terms of of the Heaviside step-function, $\Theta(\xi)$, 
as, for example in Ref.~\cite{casas}.

Just as in position space, it is possible to explicitly demonstrate
the orthonormality of the momentum-space eigenfunctions, by calculating
$\langle \phi_{m} | \phi_{n} \rangle$. Since similar methods will also
be useful in what follows, we illustrate one typical step in such an
evaluation. One required integral, for example,  is given by
\begin{equation}
{\cal I} =
\left(\frac{L}{\pi \hbar}\right)\, e^{i(n-m)\pi/2}\,
\int_{-\infty}^{+\infty}
\frac{\sin[(pL/\hbar - m\pi)/2]\,\sin[(pL/\hbar - n\pi)/2]}{(pL/\hbar - m\pi)(pL/\hbar -  n\pi)}
\, dp
\end{equation}
The denominator can be written in a form which allows use of standard
integrals, namely
\begin{equation}
\frac{1}{(pL/\hbar - m\pi)(pL/\hbar - n\pi)}
= \frac{1}{(m-n)\pi}
\left[
\frac{1}{(pL/\hbar - m\pi)}
-
\frac{1}{(pL/\hbar - n\pi)}
\right]
\, .
\label{useful_identity}
\end{equation}
Appropriate changes of variables and integrals as in
Eqn.~(\ref{important_integral}) then give simple closed form expressions,
and the complete result for $\langle \phi_m | \phi_n \rangle$
is the indeed same as in Eqn.~(\ref{orthonormality}).

\begin{figure}
\epsfig{file=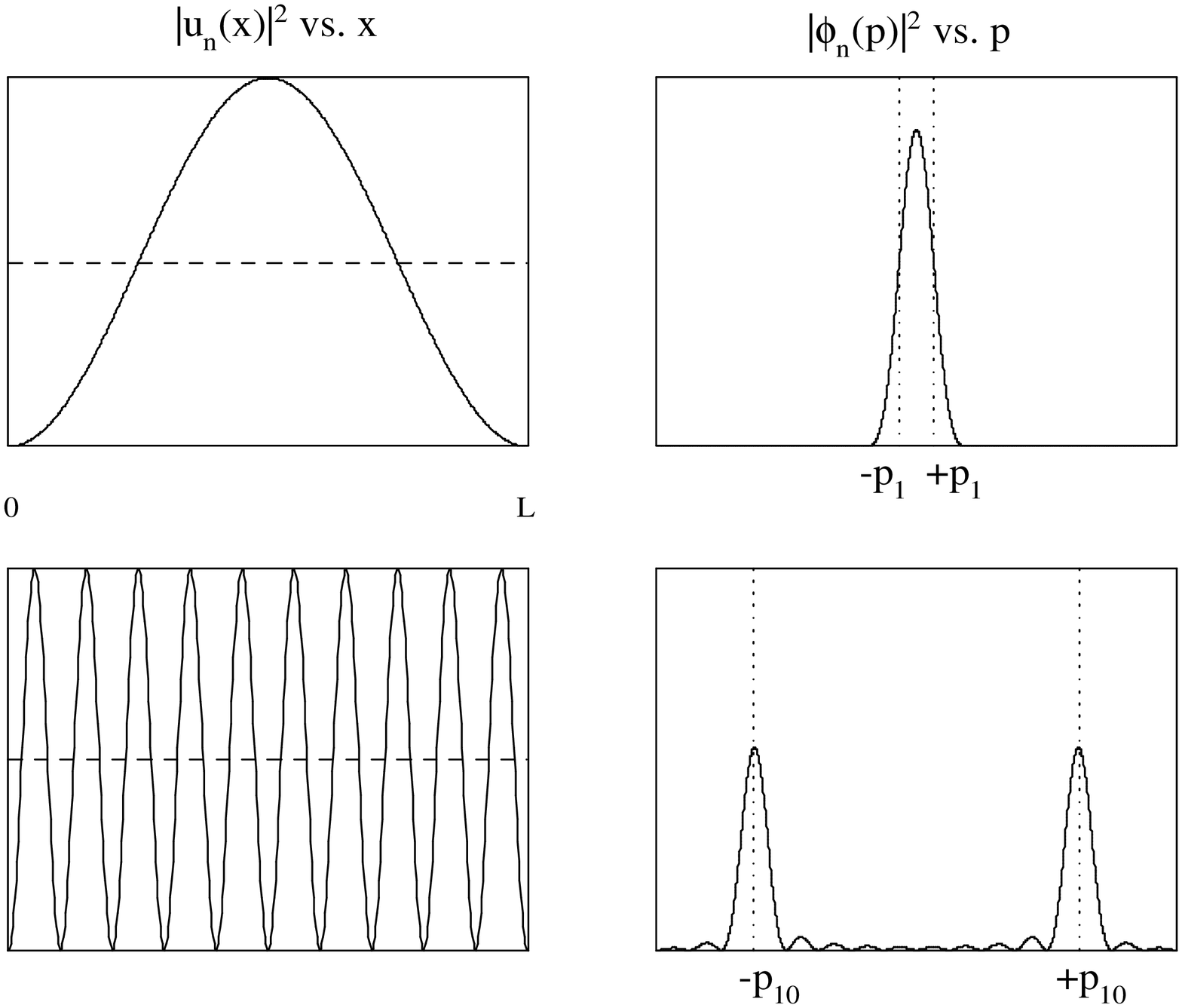,width=12cm,angle=270}
\caption{Plots of the position-space probability density,
$|u_{n}(x)|^2$ versus $x$, (left) and momentum-space probability
density,
$|\phi_{n}(p)|^2$ versus $p$, (right) for energy eigenstates in the
infinite square well for $n=1$ (top) and $n=10$ (bottom) cases. The
horizontal dashed lines on the left correspond to the (classical)
flat probability density given by $P_{CL}(x) = 1/L$
from Eqn.~(\ref{classical_position_probability}), while the vertical
dotted lines on the right correspond to the $\delta$-function classical
distribution in Eqn.~(\ref{classical_momentum_probability}).
\label{fig:xp}}
\end{figure}

Finally, for eventual comparison to the Wigner distributions for the ISW,
we plot some standard representations of the position- and momentum-space
probability densities for two low-lying states ($n=1,10$) in
Fig.~\ref{fig:xp}.

\section{Classical phase-space picture of the infinite square well}
\label{sec:phase_space_isw}

Phase-space plots of the motion of classical mechanical systems are
increasingly stressed in undergraduate textbooks on the subject,
perhaps because of their utility in the analysis or visualization of
classically chaotic systems. The correlated time dependence of $x(t)$ and
$p(t)$ in such systems is indeed easily and usefully visualized by
parametric plots in the $x-p$ plane, and the most familiar example is
perhaps the case of the harmonic oscillator. For this case, the most general
form of the solution can be written in the form
\begin{equation}
x(t) = x_A \cos(\omega t + \phi)
\qquad
\mbox{and}
\qquad
p(t) = -m\omega x_A \sin(\omega t + \phi)
\equiv p_A\sin(\omega t + \phi)
\end{equation}
and the constant value of the total energy is
\begin{equation}
\frac{[p(t)]^2}{2m} + \frac{1}{2} m \omega [x(t)]^2
= E = \frac{1}{2} m \omega^2 x_A^2
= \frac{p_A^2}{2m}
\, .
\end{equation}
This defines the elliptical phase-space path followed by the particle,
and a classical description of the joint probability densities requires that
the particle be restricted to this path, namely, that
\begin{equation}
P_{CL}(x,p)
\quad
\propto
\quad
\delta(E-p^2/2m - m\omega^2 x^2/2)
\, .
\label{classical_sho_distribution}
\end{equation}

\begin{figure}
\epsfig{file=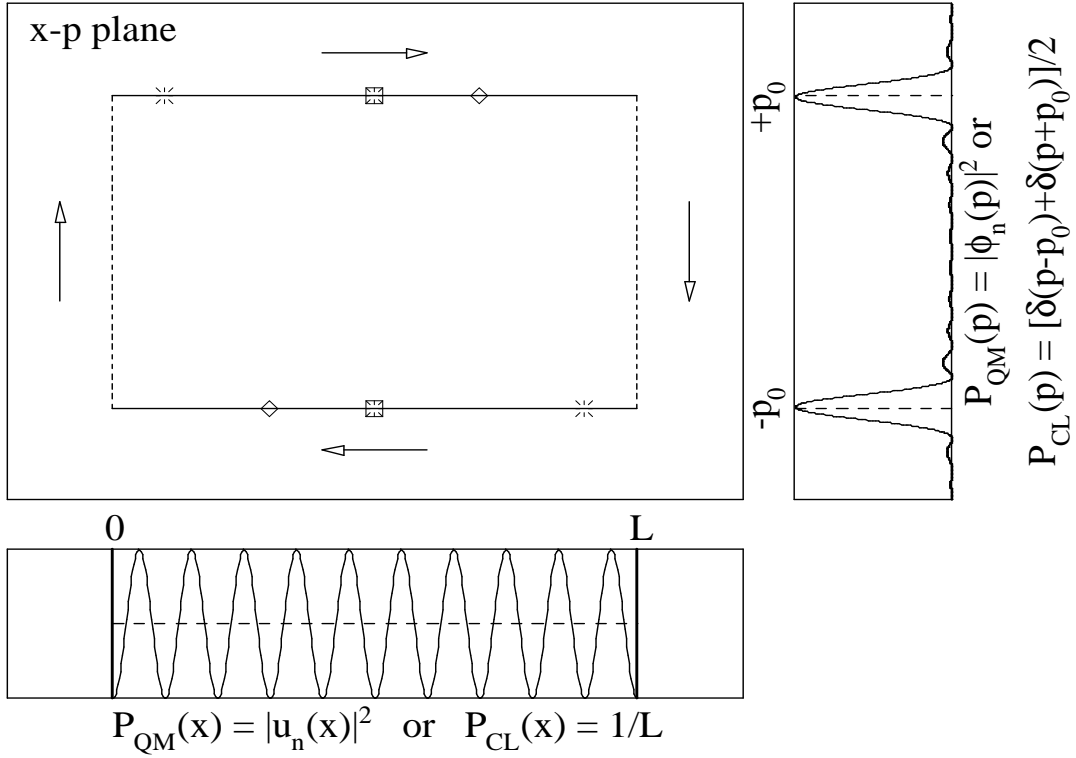,width=10cm,angle=270}
\caption{Classical phase-space picture of solutions of the infinite
square well. The classical phase-space trajectory has the particle
moving with  constant speed (momenta given by $\pm p_0$) between
the walls at $x=0,L$ (solid horizontal lines) with discontinuous changes
in velocity (momentum) due to the collisions with the walls (dashed
vertical lines). Projections onto the $x$- and $p$-axes give the
classical probability densities in
Eqns.~(\ref{classical_momentum_probability})
and
(\ref{classical_position_probability}). These are compared with the
quantum counterparts, $P_{QM}(x) = |u_{n}(x)|^2$ versus $x$, and
$P_{QM}(p) = |\phi_{n}(p)|^2$ versus $p$, for the case $n=10$.
The pairs of points in phase space indicated by squares (stars, diamonds)
are separated in time by half a classical period.
\label{fig:phase_plane}}
\end{figure}

For the classical infinite square well, the situation is at once conceptually
simpler, but somewhat more subtle mathematically. The classical paths
consist of simple `back-and-forth' motion, at constant speed
($p = \pm p_0 = \pm \sqrt{2mE}$) with sudden (discontinuous) jumps at the
walls ($x=0,L$). This is illustrated in Fig.~\ref{fig:phase_plane}
with (solid) horizontal lines at $\pm p_0$ connected by (dashed)
vertical `jumps' at the walls. The classical joint probability distribution
can then be described intuitively as
\begin{equation}
P_{CL}(x,p)
\quad
\propto
\quad
\delta(E- p^2/2m)
\qquad
\mbox{for $0<x<L$}
\end{equation}
and the corresponding classical probability densities for position,
$P_{CL}(x)$, and momentum, $P_{CL}(p)$, are obtained by integrating
over one variable or the other, as in Eqns.~(\ref{integrate_p}) and
(\ref{integrate_x}). For the $p$-space probability density, since there
is no explicit $x$ dependence inside the $\delta$-function, one finds that
\begin{equation}
P_{CL}(p) \propto  \int \delta(E-p^2/2m)\,dx
\propto \delta(E-p^2/2m)
\propto
\frac{1}{2p_0}\left[\delta(p-p_0) + \delta(p+p_0)\right]
\, .
\end{equation}
When properly normalized this reduces to
\begin{equation}
P_{CL}(p) = \frac{1}{2}\left[\delta(p-p_0) + \delta(p+p_0)\right]
\label{classical_momentum_probability}
\end{equation}
or equal probabilities of finding the particle with $p = \pm p_0$.
The corresponding position-space probability
\begin{equation}
P_{CL}(x)
\propto
\int \delta(E-p^2/2m)\,dp
\quad
\propto
\quad
\mbox{constant independent of $x$}
\end{equation}
for $0<x<L$, or when properly normalized gives
\begin{equation}
P_{CL}(x) = \frac{1}{L}
\qquad
\mbox{for $0<x<L$}
\label{classical_position_probability}
\, .
\end{equation}
These classical values are also shown in Fig.~\ref{fig:phase_plane}
(dashed horizontal line for $P_{CL}(x)$ and two vertical dashed
`spikes' for $P_{CL}(p)$) to be compared to a large quantum number
($n=10$) solution of the quantum case for comparison.

\section{Wigner distribution for the infinite well: eigenstates}
\label{sec:wigner_isw_1}

For the calculation of the Wigner distribution for energy eigenstates
in the ISW, we will first work in position space and use
\begin{equation}
P_{W}^{(n)}(x,p) = \frac{1}{\pi \hbar}
\int u_{n}(x+y)\,u_{n}(x-y)\,e^{2ipy/\hbar}\,dy
\label{wigner_eigenstate}
\end{equation}
since the position-space eigenstates, $u_n(x)$,  can be made purely real.
The limits of integration are determined by the restriction that the
$u_{n}(x \pm y)$ are non-vanishing only in the range $[0,L]$, so that we must
simultaneously satisfy the requirements
\begin{equation}
0 \leq x+y \leq L
\qquad
\mbox{and}
\qquad
0 \leq x-y \leq L
\, .
\end{equation}
This leads to upper and lower bounds for the integral over $y$ in
Eqn.~(\ref{wigner_eigenstate}) which depend on $x$ via
\begin{eqnarray}
-x \leq y \leq +x
& \qquad &
\mbox{for $0\leq x \leq L/2$} \\
-(L-x) \leq y \leq +(L-x)
& \qquad &
\mbox{for $L/2\leq x \leq L$}
\, .
\label{half_intervals}
\end{eqnarray}
Thus, over the left-half of the allowed $x$ interval, $[0,L/2]$,
we have
\begin{eqnarray}
P_{W}^{(n)}(x,p)
& = &
\frac{1}{\pi \hbar} \int_{-x}^{+x}
\left[\sqrt{\frac{2}{L}}\sin\left(\frac{n\pi(x+y)}{L}\right)\right]
\,
\left[\sqrt{\frac{2}{L}}\sin\left(\frac{n\pi(x-y)}{L}\right)\right]
\, e^{2ipy/\hbar}\, dy  \nonumber \\
& = &
\left(\frac{2}{\pi \hbar L}\right)
\left\{
\frac{\sin[2(p/\hbar - n\pi/L)x]}{4(p/\hbar - n\pi/L)}
+
\frac{\sin[2(p/\hbar + n\pi/L)x]}{4(p/\hbar + n\pi/L)}
\right.
\label{isw_wigner_distribution}
\\
& &
\qquad \qquad \qquad \qquad \qquad
\left.
-
\cos\left(\frac{2n\pi x}{L}\right)
\,
\frac{\sin(2px/\hbar)}{(2p/\hbar)}
\right\}
\nonumber
\, ,
\end{eqnarray}
while over the right-half of the interval, $[L/2,L]$, one makes the
replacement $x \rightarrow L-x$.
This form has been derived before \cite{lee}, \cite{weyl_review}, \cite{casas}
although in at least one reference it is written
in terms of Bessel functions ($j_0(z)$) which somewhat obscures its
simple derivation.
One can then demonstrate explicitly that
\begin{equation}
\int_{0}^{L} P_{W}^{(n)}(x,p)\,dx = |\phi_{n}(p)|^2
\qquad
\mbox{and}
\qquad
\int_{-\infty}^{+\infty} P_{W}^{(n)}(x,p)\,dp = |u_{n}(x)|^2
\label{project_onto_axes}
\end{equation}
using standard integrals,  or ones described in
Appendix~\ref{appendix:trig_integrals}.

The evaluation of the Wigner function using momentum-space wavefunctions,
using Eqn.~(\ref{wigner_definition_p}) integrated over
all $p$-values, is also instructive, as it naturally leads to the same
restrictions on $x$ as seen in Eqn.~(\ref{restrictions}),  as well as the
$x \rightarrow L-x$ identification over the two half-intervals, as
in Eqn.~(\ref{half_intervals}). We exhibit some of the relevant parts of
this  calculation in Appendix~\ref{appendix:wigner_momentum}.

\begin{figure}
\epsfig{file=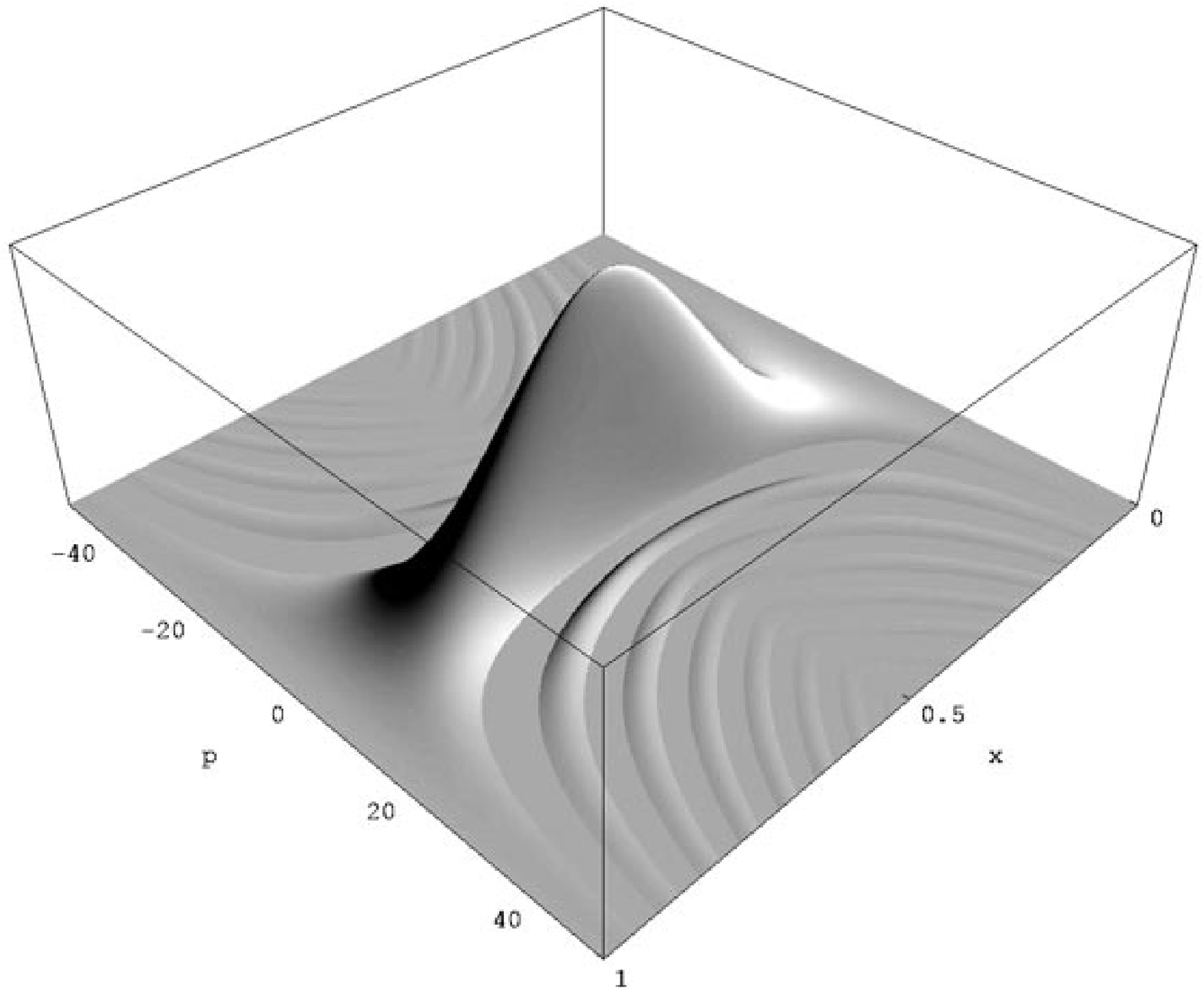,width=18cm,angle=0}
\caption{Plot of the Wigner distribution from
Eqn.~(\ref{isw_wigner_distribution}) for the
$n=1$ energy eigenstate in the infinite square well.
Only the positive ($P_{W}^{(1)}(x,p)>0$) parts are shown.
\label{fig:wigner_1}}
\end{figure}

\begin{figure}
\epsfig{file=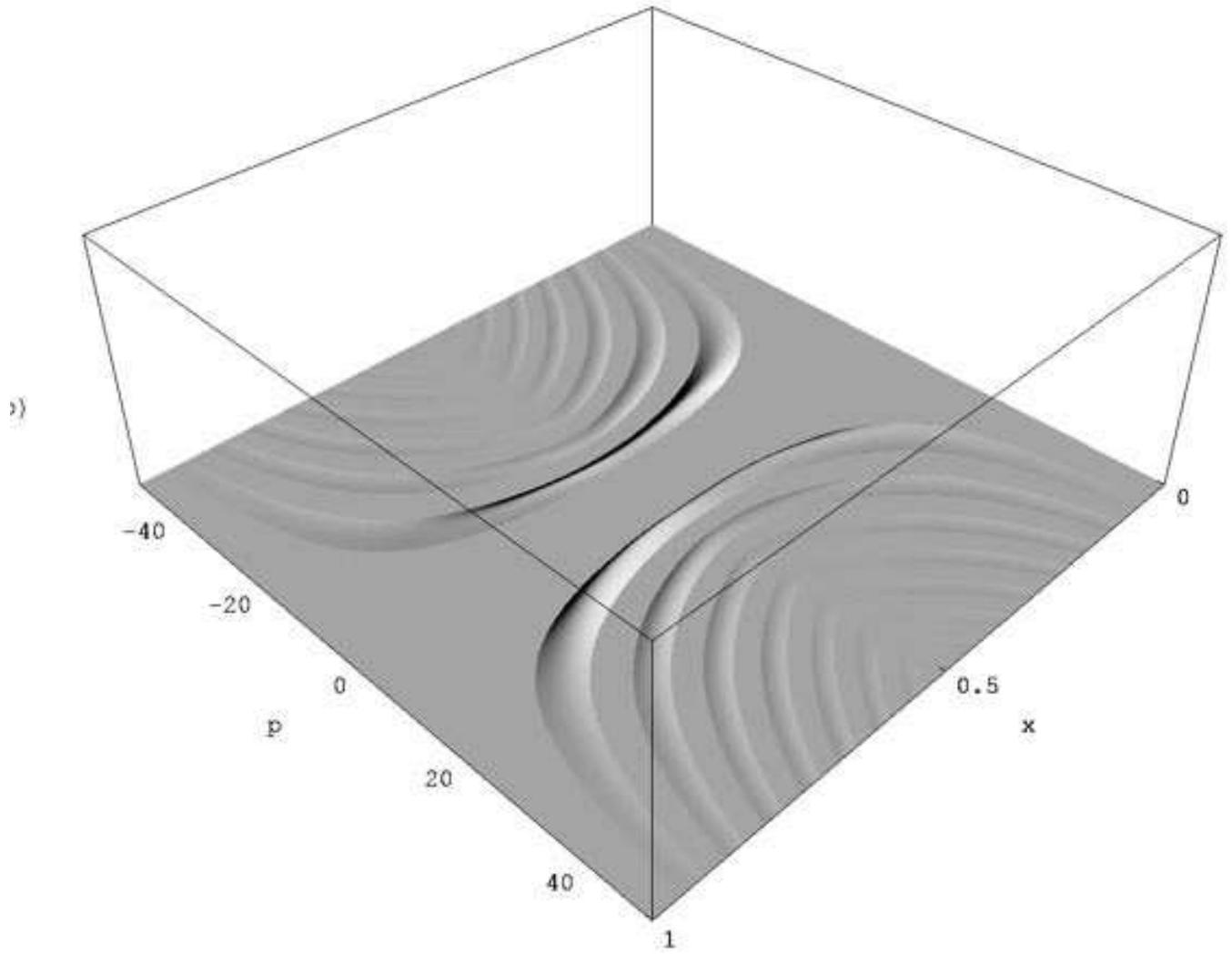,width=18cm,angle=0}
\caption{Same as Fig.~\ref{fig:wigner_1}, but with only the negative
($-P_{W}^{(1)}(x,p)>0$) parts shown. This shows that the Wigner function
for the ground state of the infinite well is almost, but not quite,
positive definite.
\label{fig:wigner_1_minus}}
\end{figure}

\begin{figure}
\epsfig{file=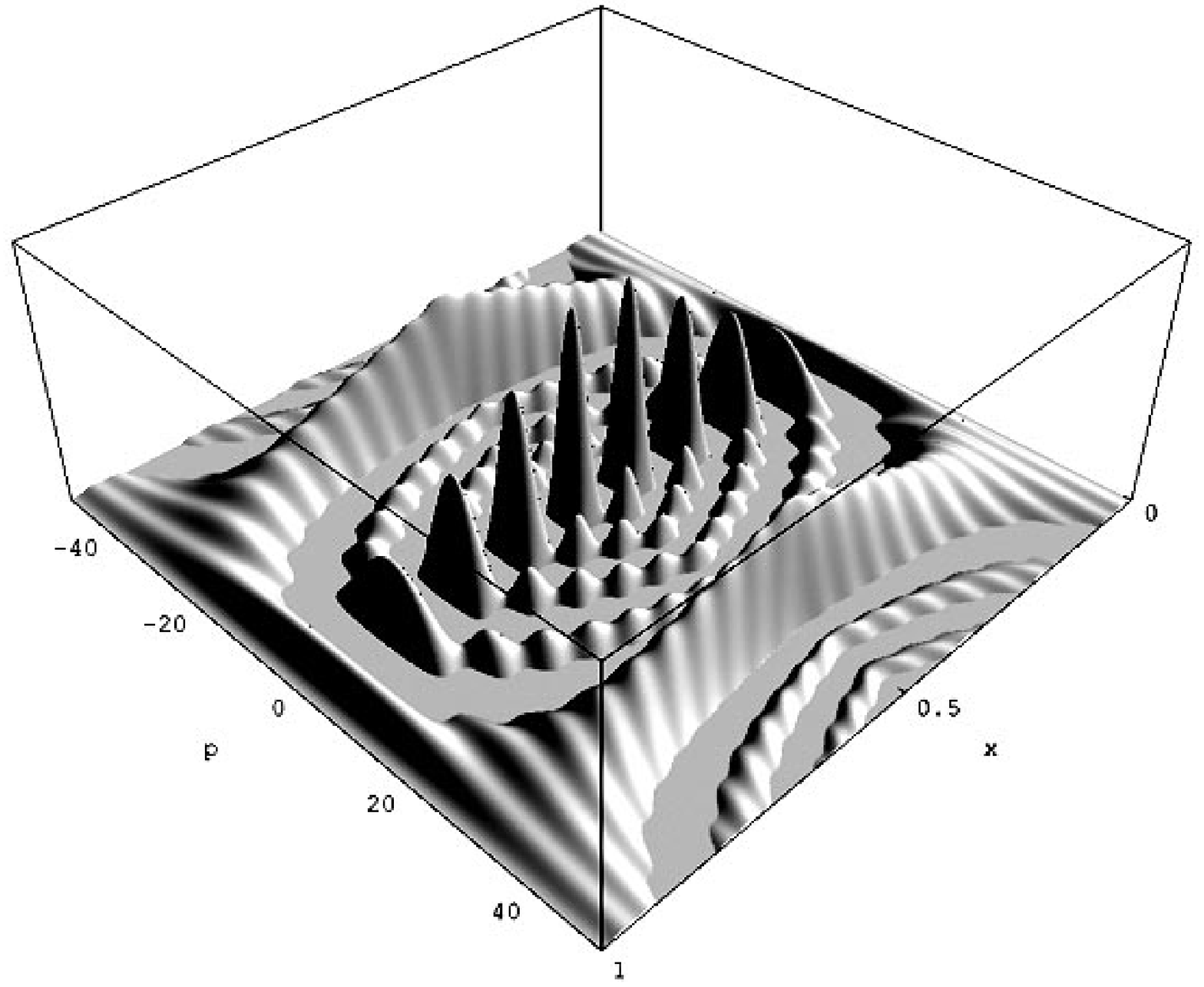,width=18cm,angle=0}
\caption{Plot of the Wigner distribution from
Eqn.~(\ref{isw_wigner_distribution}) for the
$n=10$ energy eigenstate in the infinite square well.
Only the positive ($P_{W}^{(10)}(x,p)>0$) parts are shown.
\label{fig:wigner_10}}
\end{figure}

\begin{figure}
\epsfig{file=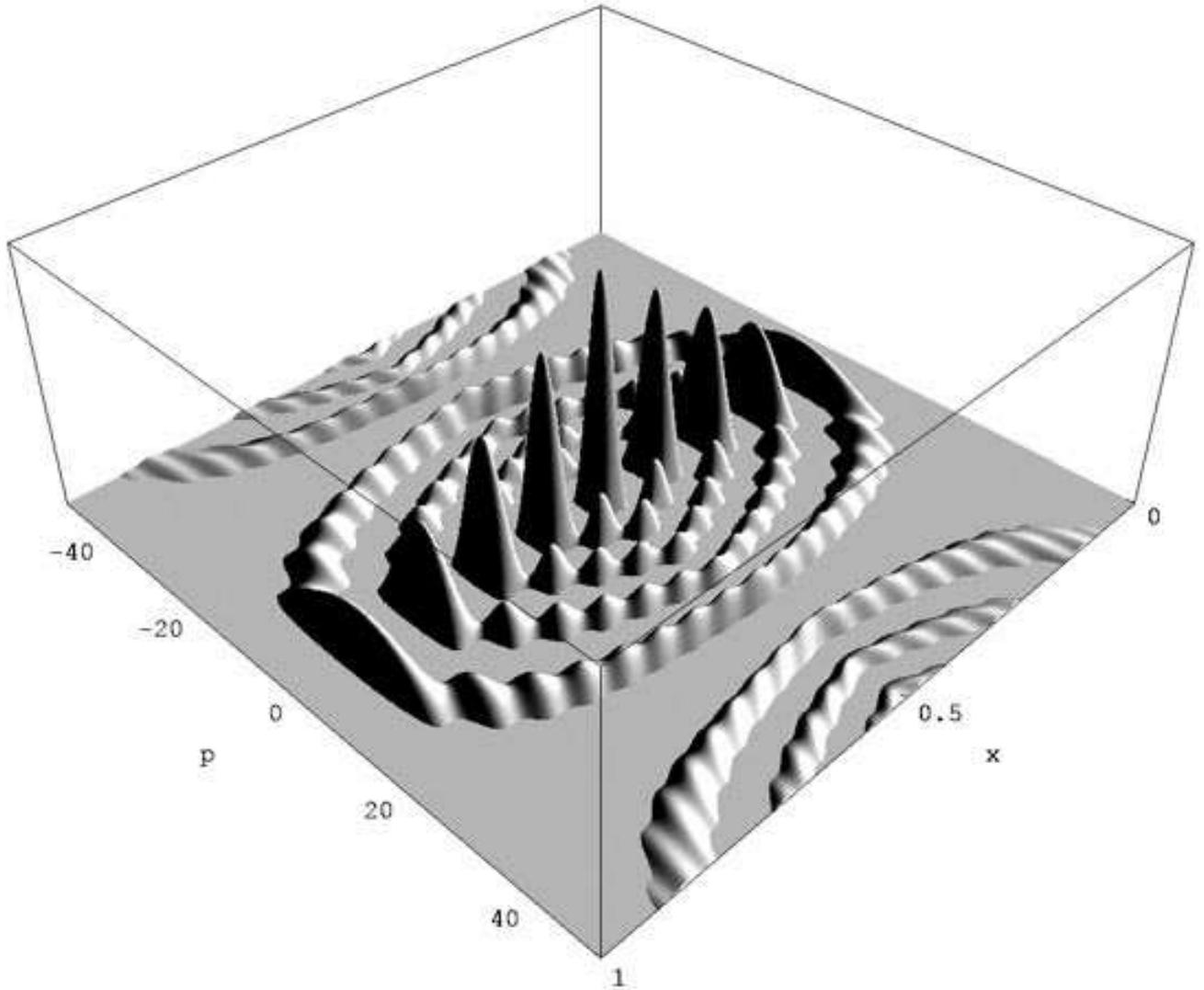,width=18cm,angle=0}
\caption{Same as Fig.~\ref{fig:wigner_10}, but with only the negative
($-P_{W}^{(10)}(x,p)>0$) parts shown.
\label{fig:wigner_10_minus}}
\end{figure}

With these results in hand, we exhibit examples of the Wigner function
for the ISW, for the $n=1$ ground state (Figs.~\ref{fig:wigner_1}
and \ref{fig:wigner_1_minus}) and for an excited ($n=10$) state
(Figs.~\ref{fig:wigner_10} and \ref{fig:wigner_10_minus}), using $\hbar=L=1$
for simplicity. For ease of visualization, for both cases we show separate
plots for $P_{W}^{(n)}(x,p)>0$ and $P_{W}^{(n)}(x,p)<0$ values.
For the ground state, the projection of $P^{(n)}_{W}(x,p)$ onto the
$x$- and $p$-axes to obtain $|u_{n}(x)|^2$ and $|\phi_{n}(p)|^2$ (as in
Fig.~\ref{fig:xp}) via Eqn.~(\ref{project_onto_axes}) is straightforward
to visualize, and we also note that this relatively smooth ground state
wavefunction gives a Wigner distribution which is almost everywhere
positive (Fig.~\ref{fig:wigner_1}), but with a small negative contribution
as seen most clearly in Fig.~\ref{fig:wigner_1_minus}.

For the $n=10$ (or any $n>>1$) case, the structures are more surprising,
but still yield the appropriate probability densities upon projection onto
the $x$- and $p$-axes. The obvious `triangular' form of both the `fin'-shaped
features along the $p = \pm p_n$ axes and the central `spines' along the
$p=0$ axes can be easily seen to arise from the form in
Eqn.~(\ref{isw_wigner_distribution}) in those limits. For those
cases, the $\sin[2Fx/\hbar]/F$ form (where $F= (p-p_n)$, $(p+p_n)$, or $p$)
of each term (for $0<x<L/2$) gives a linear
dependence on $x$ when $F\rightarrow 0$, while for $L/2<x<L$ this is
replaced by the $L-x$ factor, giving the triangular dependence obvious
from Fig.~\ref{fig:wigner_10}. The  smooth features along the $p = \pm p_n$
axes (seen only for positive values of $P_{W}^{(n)}(x,p)$) are obviously
reminiscent of the classical phase-space picture in
Fig.~\ref{fig:phase_plane}, while the highly oscillatory structures
`inside' the classical rectangular boundary (obvious in both
Figs.~\ref{fig:wigner_10} and \ref{fig:wigner_10_minus}) locally average to
the small, but non-vanishing values of the cross-term in
Eqn.~(\ref{momentum_space_probability}) and are just as clearly purely
quantum mechanical in origin.

\section{Wigner distribution for the infinite well: wave packet solutions}
\label{sec:wigner_isw_2}

For a general wave packet solution for the infinite square well, we require
the on- and off-diagonal terms in Eqn.~(\ref{off_diagonal}). Using the
position-space eigenstates in Eqn.~(\ref{1d_eigenfunctions}), over the
interval $[0,L/2]$ we find that
\begin{eqnarray}
P_{W}^{(m,n)}(x,p) & = &
\frac{1}{\pi \hbar} \int_{-x}^{+x}
\left[\sqrt{\frac{2}{L}}\sin\left(\frac{m\pi(x+y)}{L}\right)\right]
\,
\left[\sqrt{\frac{2}{L}}\sin\left(\frac{n\pi(x-y)}{L}\right)\right]
\, e^{2ipy/\hbar}\, dy  \nonumber \\
& = &
\frac{1}{\pi \hbar}
\left[
e^{+i(m-n)\pi x/L}\,\frac{\sin[(2p/\hbar +(m+n)\pi/L)x]}{2pL/\hbar + (m+n)\pi}
\right. \nonumber \\
& &
\qquad \qquad \qquad
\left.
+
e^{-i(m-n)\pi x/L}\,\frac{\sin[(2p/\hbar -(m+n)\pi/L)x]}{2pL/\hbar - (m+n)\pi}
\right. \\
\label{off_diagonal_wigner}
& &
\qquad \qquad \qquad
\left.
-
e^{+i(m+n)\pi x/L}\,\frac{\sin[(2p/\hbar+(m-n)\pi/L)x]}{2pL/\hbar + (m-n)\pi}
\right. \nonumber \\
& &
\qquad \qquad \qquad
\left.
-
e^{-i(m+n)\pi x/L}\,\frac{\sin[(2p/\hbar -(m-n)\pi/L)x]}{2pL/\hbar - (m-n)\pi}
\right] \nonumber
\end{eqnarray}
and it is easy to check that this result reduces to the expression
in Eqn.~(\ref{isw_wigner_distribution}) when $m=n$. In order to extend this
to the interval $[L/2,L]$,  it is important to note that the substitution
$x \rightarrow L-x$ should be made {\it only} in those terms arising from
the integration over $dy$, namely, the $\sin[(2p/\hbar \pm (m\pm n)/L)x]$
terms.

These results can then be used to evaluate the time-dependent Wigner
function for any initial wave packet in the infinite square well, with
the simplest non-trivial example being that for a two-state system,
$\psi(x,0) = \gamma u_{1}(x) + \delta u_{2}(x)$, which can then
be compared to more standard representations in position space or
momentum space \cite{robinett_book}. Such states, while exhibiting non-trivial
dynamical behavior (sometimes said to mimic time-dependent radiating
systems \cite{dean}),  are still highly quantum mechanical and to approach the
semi-classical limit, we require more localized states, constructed
from higher quantum number eigenfunctions.

We can examine the time dependence of such an initially localized state by
choosing a standard Gaussian of the form
\begin{equation}
\psi_{(G)}(x,0)
=
\frac{1}{\sqrt{b\sqrt{\pi}}}
\,
e^{-(x-x_0)^2/2b^2}
\,
e^{ip_0(x-x_0)/\hbar}
\end{equation}
where we will always assume that $x_0$ is chosen such that
$\psi_{(G)}(x,0)$ is sufficiently contained within the well so that we make
an exponentially small error by neglecting any overlap with the region
outside the walls, and may thus also ignore any problems associated with
possible discontinuities at the wall. In practice, this condition
only requires the wave packet to be a few times
$\Delta x_0 = b/\sqrt{2}$ away from an infinite wall boundary.
With these assumptions, we can then extend the integration region from the
finite $[0,L]$ interval to the entire 1-D space, giving the (exponentially
good) approximation for the expansion coefficients \cite{born},
\cite{blueprint}
\begin{eqnarray}
a_{n} & = & \int_{0}^{L}
\left[u_{n}(x)\right]\, \left[\psi_{G}(x,0)\right]\,dx \nonumber \\
& \approx & \int_{-\infty}^{+\infty}
\left[u_{n}(x)\right]\, \left[\psi_{G}(x,0)\right]\,dx
\\
& = &
\left(\frac{1}{2i}\right)
\sqrt{\frac{4b\pi}{L\sqrt{\pi}}}
[
e^{in \pi x_0/L}  e^{-b^2(p_0 + n\pi \hbar/L)^2/2\hbar^2}
-
e^{-in \pi x_0/L} e^{-b^2(p_0 - n\pi \hbar/L)^2/2 \hbar^2}]
\, .
\nonumber
\label{approximate_expansion}
\end{eqnarray}
The position-space and momentum-space wavefunctions are then given by
\begin{equation}
\psi(x,t) = \sum_{n=1}^{\infty} a_n u_n(x)\,e^{-iE_nt/\hbar}
\qquad
\mbox{and}
\qquad
\phi(p,t) = \sum_{n=1}^{\infty} a_n \phi_n(p)\,e^{-iE_nt/\hbar}
\end{equation}
while the time-dependent Wigner distribution is given by
Eqn.~(\ref{time_dependent_wigner}).

The time dependence of any general quantum state (not necessarily the ISW)
is determined by the $\exp(-iE_nt/\hbar)$ factors, and for highly localized,
semi-classical wave packets, where one typically can expand $E(n)$ about
a (large) central value of the quantum number,  $n_0>>1$, one can write
this dependence in the form \cite{bluhm_ajp}, \cite{robinett_physics_report}
\begin{eqnarray}
e^{-iE_nt/\hbar} & = & \exp\left( -i/\hbar\left[E(n_0)t + (n-n_0) E'(n_0)t +
\frac{1}{2} (n-n_0)^2 E''(n_0) t \right. \right. \nonumber \\
& &
\qquad \qquad \qquad
\left. \left.
+ \frac{1}{6}(n-n_0)^3 E'''(n_0)t  + \cdots
\right]\right) \nonumber \\
& \equiv  & \exp \left( -i\omega_0 t - 2\pi i(n-n_0) t/T_{cl}
- 2\pi i(n-n_0)^2t/T_{rev} \right.
\label{expansion_in_time}\\
& & \qquad \qquad \qquad \left. - 2\pi i(n-n_0)^3t/T_{super} + \cdots \right)
\nonumber
\end{eqnarray}
where each term in the expansion (after the first) defines an important
characteristic time scale, via
\begin{equation}
T_{cl} = \frac{2\pi \hbar}{|E'(n_0)|}
,
\quad
%,
\quad
T_{rev} = \frac{2\pi \hbar}{|E''(n_0)|/2}
,
\quad
%,
\quad
\mbox{and}
\quad
T_{super} = \frac{2\pi \hbar}{|E'''(n_0)|/6}
\, ,
\label{all_relevant_times}
\end{equation}
namely, the classical period ($T_{cl}$), the revival time ($T_{rev}$),
and the super-revival time ($T_{super}$).
The classical periodicity for the ISW in this formalism is given by
\begin{equation}
T_{cl} = \frac{2\pi\hbar}{|E'(n_0)|}
= \frac{2mL^2}{\hbar \pi n_0}
= \frac{2L}{[(\hbar \pi n_0/L)/m]}
= \frac{2L}{v_{n_0}}
\qquad
\quad
\mbox{where}
\quad
\qquad
v_{n_0} \equiv \frac{p_{n_0}}{m}
\label{infinite_well_period}
\end{equation}
and $v_{n_0}$ is the analog of the classical speed, giving a result
in agreement with classical expectations. Depending on the relative values
of $T_{cl}$ and the spreading time, $t_0$, there can be many obvious
similarities to the classical motion, easily visualized through an
expectation value analysis \cite{robinett_expectation}.

As an example of quasi-classical time evolution of a quantum state,
we consider a Gaussian wave packet characterized by the parameters
\begin{eqnarray}
2m = L = \hbar = 1
\qquad
& &
\qquad
\Delta x_0 = \frac{b}{\sqrt{2}} = \frac{1}{20} << 1 = L
\\
p_0 = +\frac{40\pi\hbar}{L} = 40\pi >> 10 = \Delta p_0
= \frac{\hbar}{2\Delta x_0}
\qquad
& &
\qquad
x_0 = L/2 = 0.5
\, .
\label{gaussian_parameters}
\end{eqnarray}
We use a central values of $n_0 = 40$ so that the ratio of classical
periodicity to spreading time is $T_{cl}/t_0 = 10/\pi$ and significant
spreading can be seen even over half a classical period.

\begin{figure}
\epsfig{file=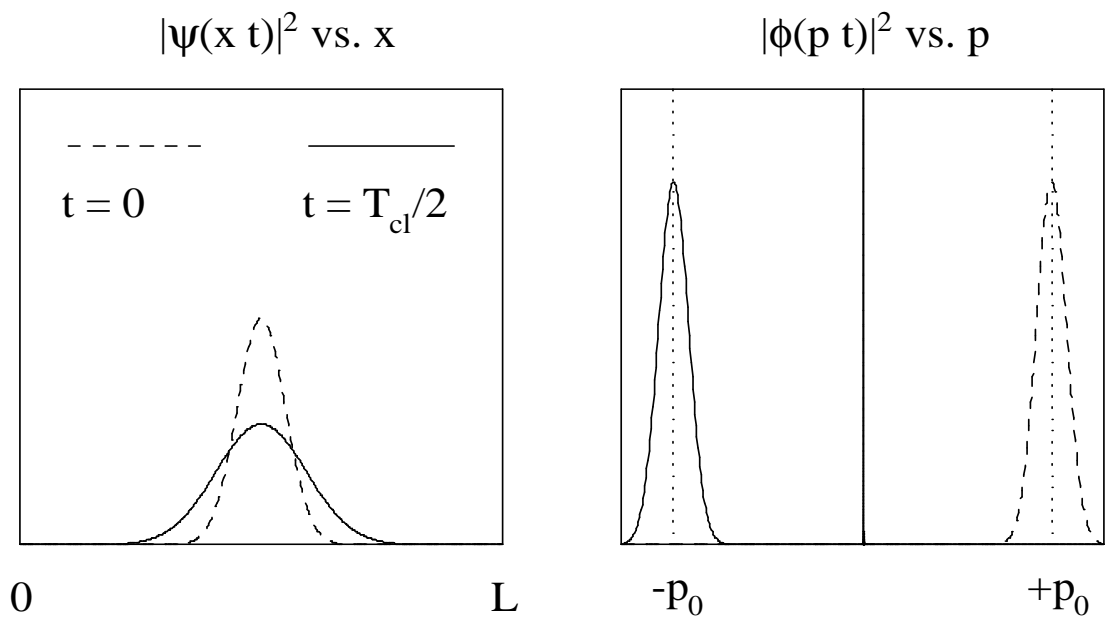,width=8cm,angle=270}
\caption{Plots of the position-space probability density,
 $|\psi(x,t)|^2$ versus $x$, and momentum-space probability
density, $|\phi(p,t)|^2$ versus $p$, for a Gaussian wave packet
solution in the ISW. Times corresponding to $t=0$ (dashed) and
$T_{cl}/2$ (solid) are shown. The parameters of
Eqn.~(\ref{gaussian_parameters}) and $n_0 = 40$ are used.
\label{fig:xp_early}}
\end{figure}

\begin{figure}
\epsfig{file=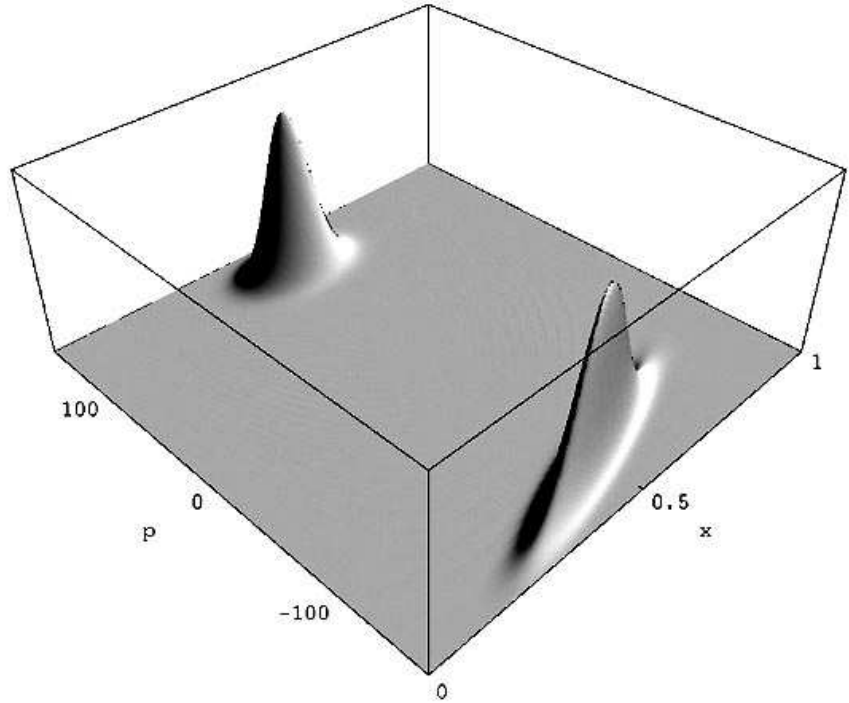,width=12cm,angle=0}
\caption{Plot of the Wigner function, $P_{W}(x,p;t)$ versus
$(x,p)$ as a function of time for $t=0$ and $t=T_{cl}/2$,
to be compared to Fig.~\ref{fig:xp_early}.
\label{fig:wigner_early}}
\end{figure}

We first show in Fig.~\ref{fig:xp_early} the position- and
momentum-space probability densities for the initial wave packet
($t=0$, dashed) and half a classical period later ($t=T_{cl}/2$,
solid). The spreading in position space is clearly visible, while
the switch from momentum values centered around $+p_0$ to $-p_0$
after the first `collision' with the infinite wall is also
apparent. For comparison, we show  $P_{W}(x,p;t)$ for the same two
times in Fig.~\ref{fig:wigner_early} and note the two highly
localized `lumps' of probability, centered at the appropriate
locations in phase space, but with the obvious spreading in
position space, consistent with the form in
Eqn.~(\ref{free_particle_gaussian_wigner}), and with positive-definite
values in each case. This behavior is
similar to the classical phase space picture in
Fig.~\ref{fig:phase_plane} for the two states denoted by squares,
separated by $T_{cl}/2$.

\begin{figure}
\epsfig{file=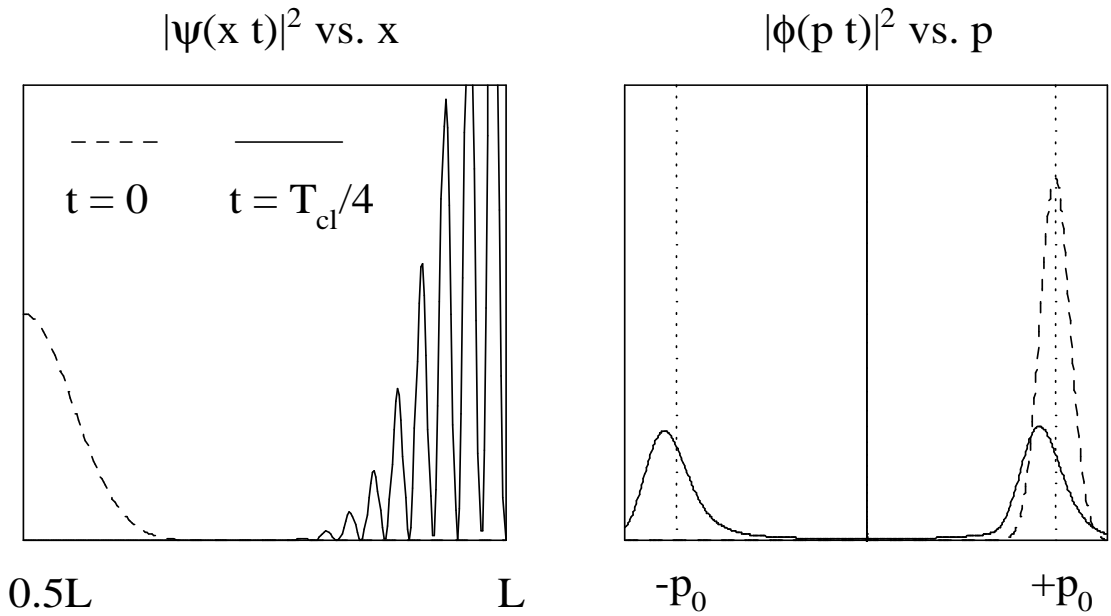,width=8cm,angle=270}
\caption{Same as Fig.~\ref{fig:xp_early}, except for $t=0$ (dashed)
and $t=T_{cl}/4$ where the classical particle would be hitting the wall.
(Recall that the two momentum peaks for the `collision' time are not
symmetrically placed at $\pm p_0$ \cite{doncheski_splash} since the
high-momentum components of the initial wave packet arrive at the wall,
and hence are also reflected, first.)  Note that the $|\psi(x,t)|^2$ is
plotted over the interval $[L/2,L]$ to show the collision with the
wall in more detail.
\label{fig:xp_splash}}
\end{figure}

\begin{figure}
\epsfig{file=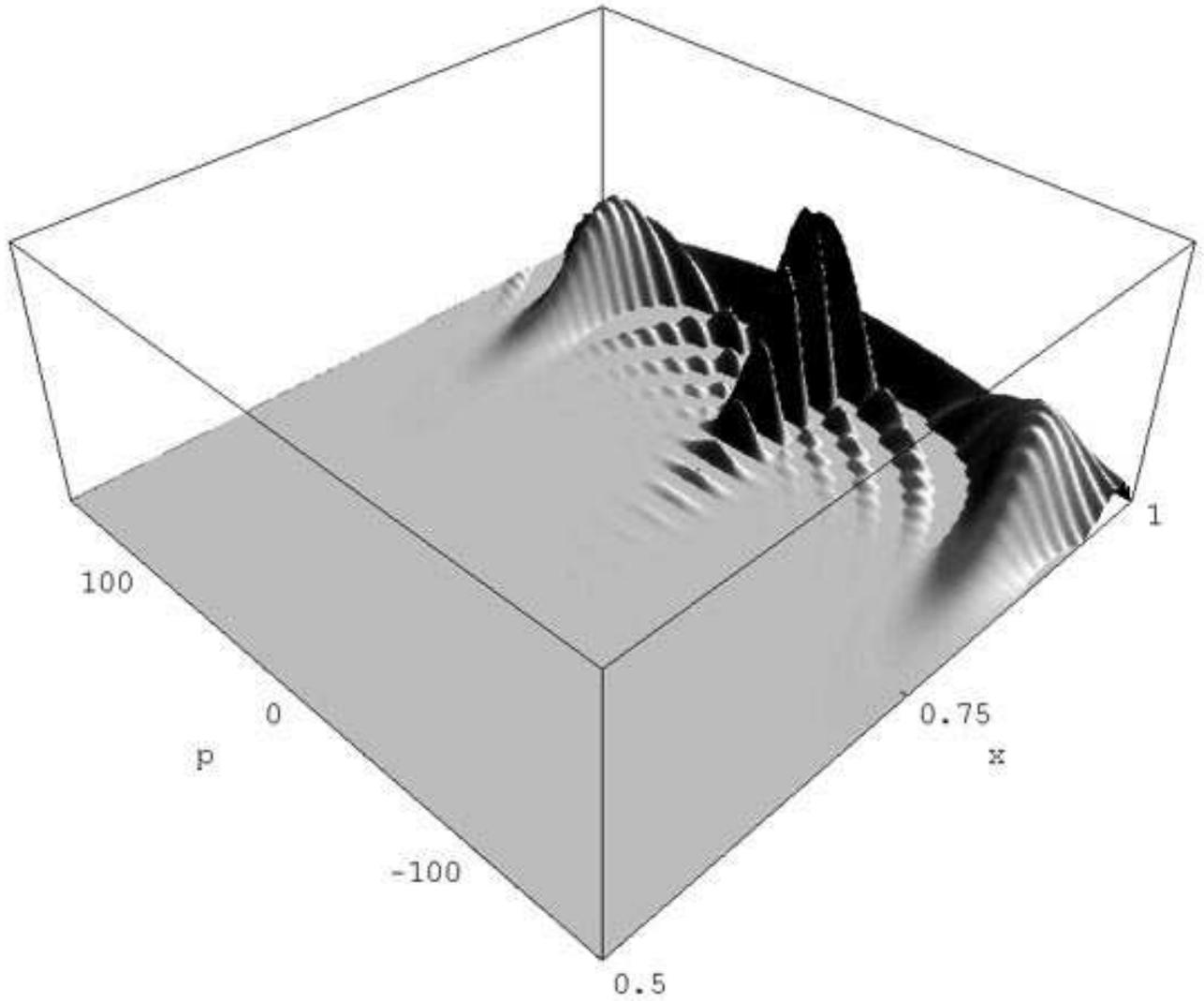,width=18cm,angle=0}
\caption{Same as Fig.~\ref{fig:wigner_early}, but for $t= T_{cl}/4$
where the classical particle would be hitting the wall. Only positive 
values ($P_{W}(x,p;t)>0$) are shown and the $x$ interval $[L/2,L]$ is
used, as in Fig.~\ref{fig:xp_splash}. 
\label{fig:wigner_splash}}
\end{figure}

For comparison to this quasi-classical behavior, we also show results
for $t=T_{cl}/4$, corresponding to the time when the classical particle
would `hit' the wall, both for the position- and momentum-space probability
densities (in Fig.~\ref{fig:xp_splash}) as well as for the Wigner
function (in Fig.~\ref{fig:wigner_splash}). In this case, the
description of the time-dependent solution in terms of sums of image
states \cite{born},  \cite{kleber_exact}, \cite{andrews} is appropriate,
with the large, interference term at $p \approx 0$
in the Wigner distribution arising from effects such as in
Eqn.~(\ref{two_gaussian_wigner}). As discussed above, the Wigner function 
is positive-definite only at times when it can be well approximated 
as a single isolated Gaussian, such as at $t=0$ and $t=T_{cl}/2$ in 
Fig.~\ref{fig:wigner_early}, or, as we will see, at the revival time
$T_{rev}$. As seen here at $T_{cl}/4$, at many other times the Wigner
function can become negative and we will henceforth only show the 
$P_{W}(x,p;t)>0$ values for ease of visualization.

For longer time scales, we require the revival time, $T_{rev}$, which is
given by
\begin{equation}
T_{rev} \equiv \frac{2\pi \hbar}{|E''(n_0)|/2}
= \frac{2\pi \hbar}{E_0}
= \frac{4mL^2}{\hbar \pi}
= (2n_0) T_{cl}
\label{infinite_well_revival}
\end{equation}
and for the infinite square well no longer time scales are present due
to the purely quadratic dependence of the energy eigenvalues. The revival time
scale can clearly be much larger than the classical period for wave packets
characterized by $n_0>>1$ (giving $T_{rev}/T_{cl} = 80$ for the examples
presented here.)

The quantum revivals for this system are exact since we have
\begin{equation}
\psi(x,t+T_{rev}) =
\sum_{n=1}^{+\infty}
a_n u_{n}(x) e^{-iE_n(t+T_{rev})/\hbar}
=
\sum_{n=1}^{+\infty}
a_n u_{n}(x) e^{-iE_n t/\hbar}
e^{-i2\pi n^2}
= \psi(x,t)
\end{equation}
and any wave packet returns to its initial state after a time $T_{rev}$.
At half this time, $t=T_{rev}/2$, the wave packet also 
reforms \cite{aronstein}, since
\begin{eqnarray}
\psi(L-x,t+T_{rev}/2)
& = &
\sum_{n=1}^{\infty}
a_{n} u_{n}(L-x) e^{-iE_n t/\hbar} e^{-iE_n T_{rev}/2\hbar}
\nonumber \\
& = &
\sum_{n=1}^{\infty} a_n u_{n}(x) e^{-iE_n t/\hbar}\left[ (-1)^{n+1} e^{-in^2\pi}\right]
\label{mirror_expression}\\
& = &
-\psi(x,t)
\nonumber
\end{eqnarray}
where we have used the symmetry properties of the $u_{n}(x)$ from
Eqn.~(\ref{general_parity}). This implies that
\begin{equation}
|\psi(x,t+T_{rev}/2)|^2 = |\psi(L-x,t)|^2
\label{infinite_mirror_position}
\end{equation}
so that at half a revival time later, any initial wave packet will
reform itself (same shape, width, etc.), but at a location
mirrored about the center of the well. Using 
Eqn.~(\ref{mirror_expression}), one can also show that
\begin{equation}
|\phi(p,t+T_{rev}/2)|^2 = |\phi(-p,t)|^2
\end{equation}
so that the packet is moving with `mirror' (opposite) momentum values 
as well, and therefore in the opposite corner of phase space.

\begin{figure}
\epsfig{file=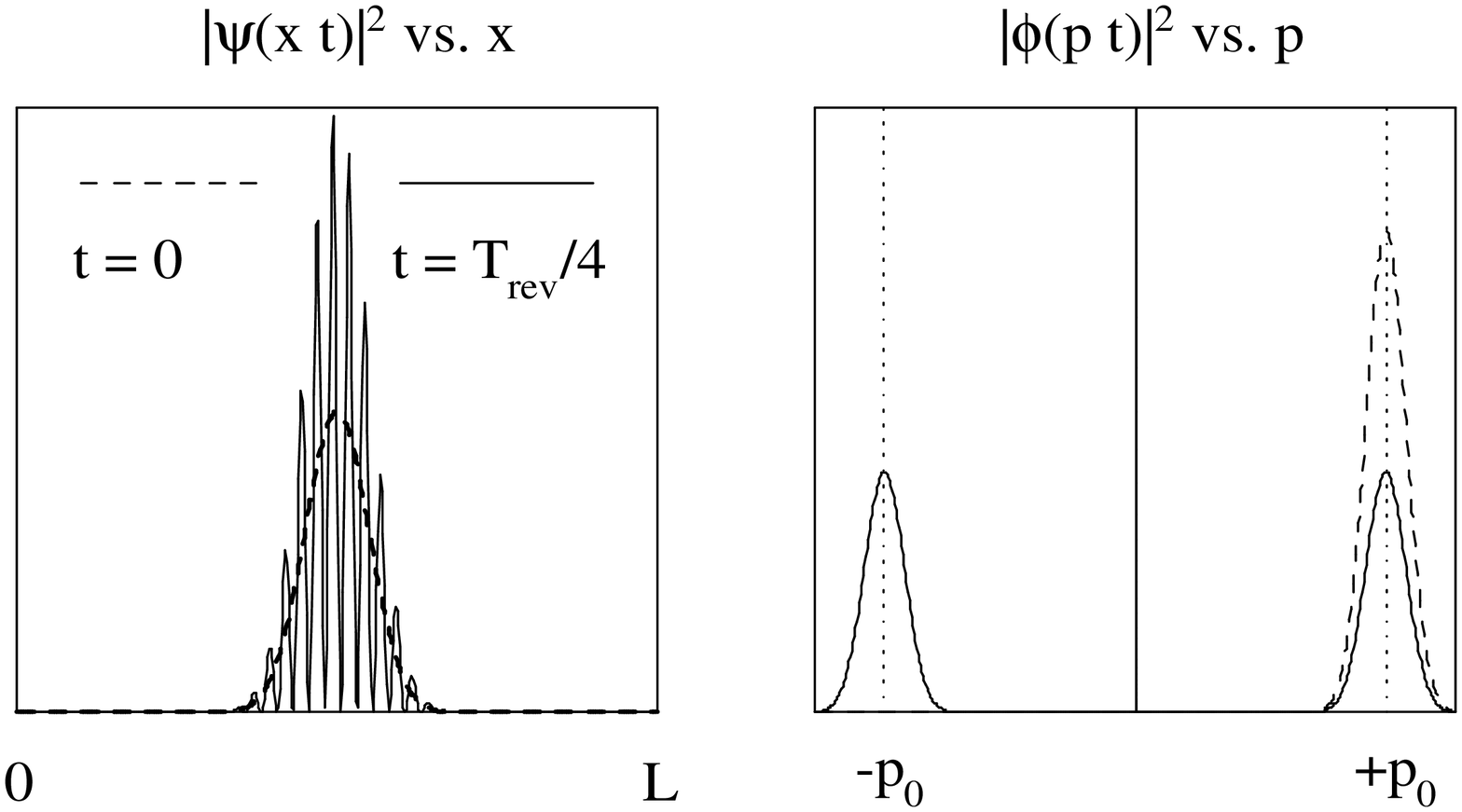,width=8cm,angle=270}
\caption{Same as Fig.~\ref{fig:xp_early}, except for $t=0$ (dashed)
and for a fractional revival at $t = T_{rev}/4$ (dashed)
\label{fig:xp_quarter}}
\end{figure}

\begin{figure}
\epsfig{file=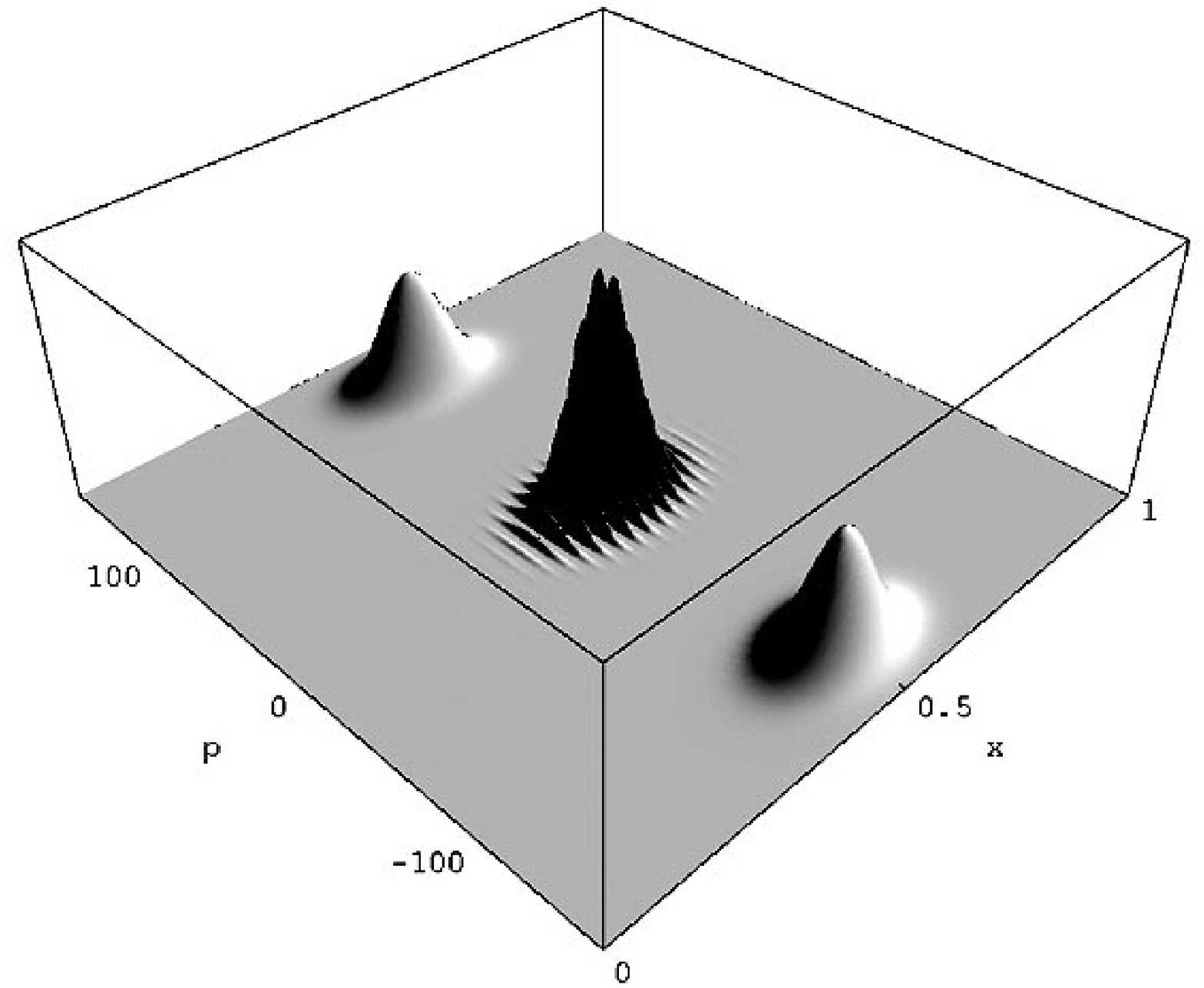,width=18cm,angle=0}
\caption{Same as Fig.~\ref{fig:wigner_early}, but for
the fractional revival at $t=  T_{rev}/4$, to be compared to
Fig.~\ref{fig:xp_quarter}. 
Only positive values ($P_{W}(x,p;t)>0$) are shown.
\label{fig:wigner_quarter}}
\end{figure}

More interestingly, at various fractional multiples of the revival
time, $pT_{rev}/q$, the wave packet can also reform as several
small copies (sometimes called `mini-packets' or `clones') of the 
original wave packet, with well-defined phase
relationships. (The original mathematical arguments showing how this
behavior arises in general wave packet solutions were made by Averbukh
and Perelman in Ref.~\cite{averbukh}). For example, near
$t \approx T_{rev}/4$ the wave packet can be written as a linear
combination of the form
\begin{equation}
\psi(x,t\approx T_{rev}/4)
=
\frac{1}{\sqrt{2}}
\left[
e^{-i\pi/4}\, \psi_{cl}(x,t) +
e^{+i\pi/4}\, \psi_{cl}(x,t+T_{cl}/2)
\right] \nonumber
\end{equation}
where $\psi_{cl}(x,t)$ describes the short-term time development
of the initial wave packet obtained by keeping only the terms linear
in $n$ in the expansion in Eqn.~(\ref{expansion_in_time}).
The position- and momentum-space probability densities at $T_{rev}/4$ are
shown in Fig.~\ref{fig:xp_quarter} illustrating this phenomena, but this
behavior is much more interestingly visualized through the Wigner
function description in Fig.~\ref{fig:wigner_quarter}. The Wigner function
in this case is well-represented by the terms in
Eqn.~(\ref{two_gaussian_wigner}), where the cross-term is only oscillatory
in the $p$ variable, since the corresponding values of
$x_A = x_B = L/2 = x_0$ are the same in this case.

\begin{figure}
\epsfig{file=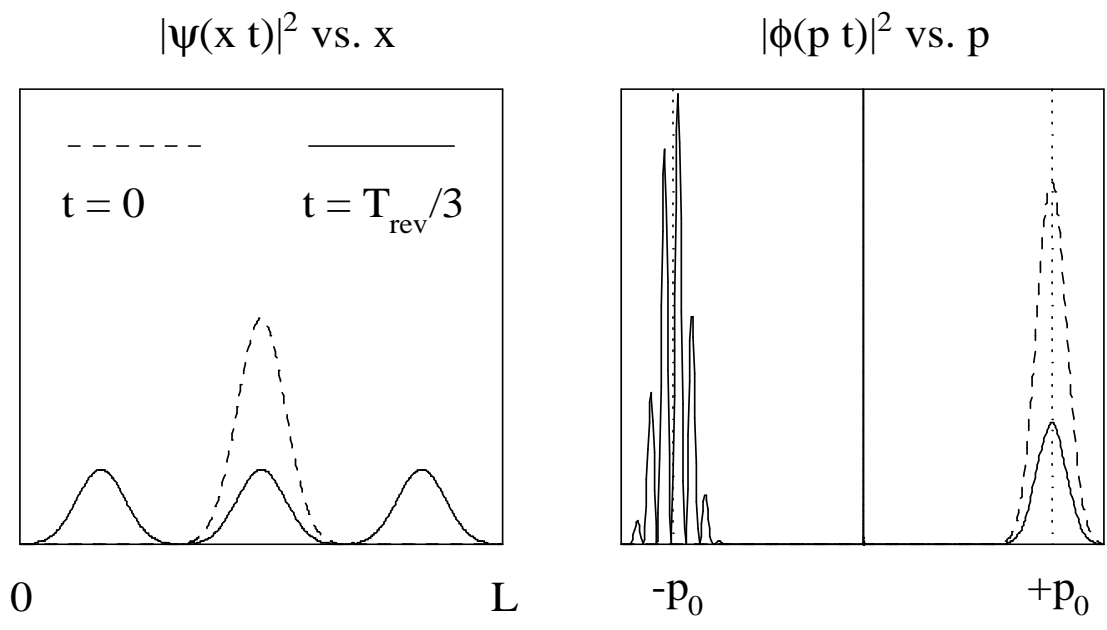,width=8cm,angle=270}
\caption{Same as Fig.~\ref{fig:xp_early}, except for $t=0$ (dashed)
and for a fractional revival at $t = T_{rev}/3$ (solid).
\label{fig:xp_third}}
\end{figure}

\begin{figure}
\epsfig{file=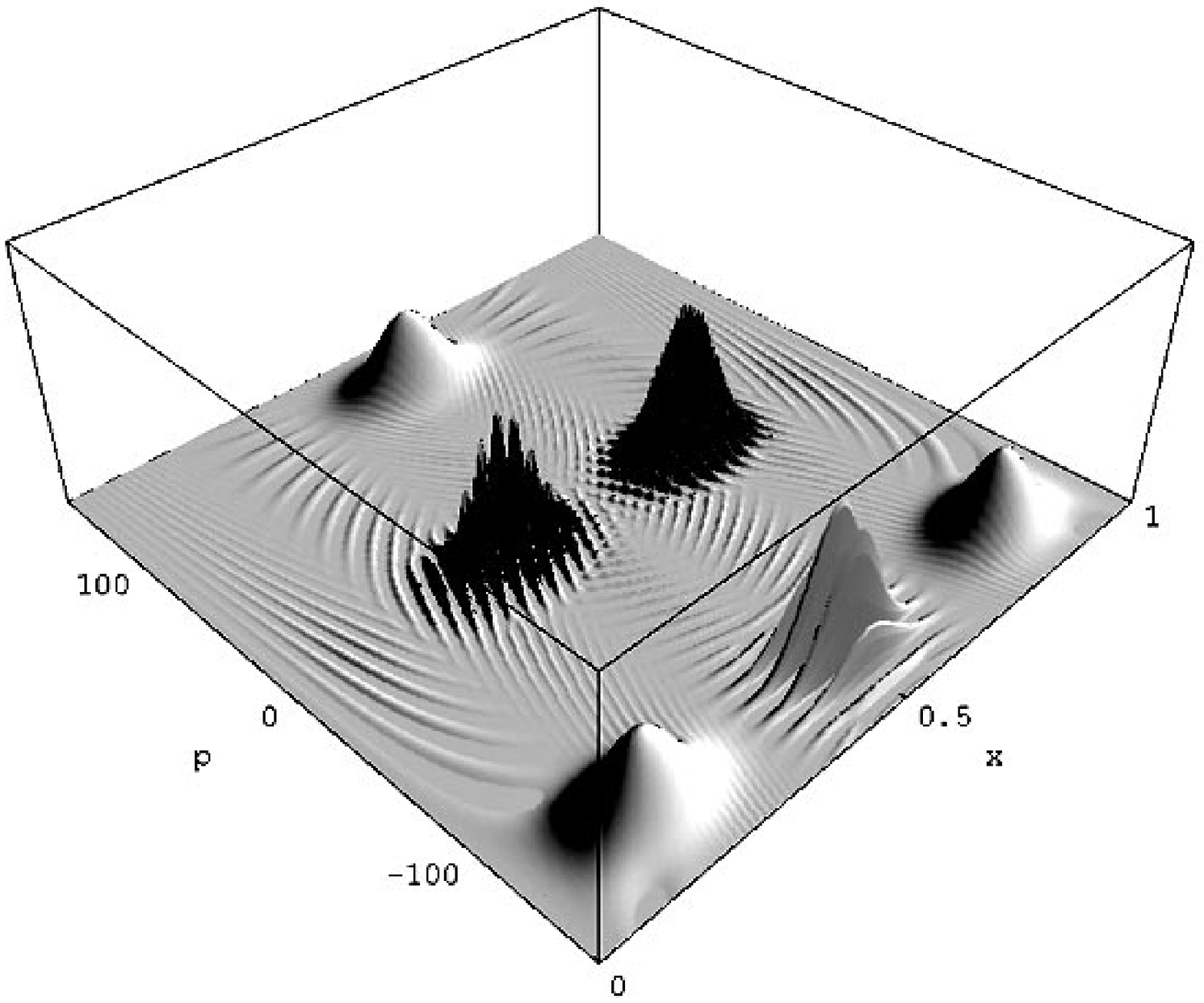,width=18cm,angle=0}
\caption{Same as Fig.~\ref{fig:wigner_early}, but for
the fractional revival at $t=  T_{rev}/3$, to be compared to
Fig.~\ref{fig:xp_third}.
Only positive values ($P_{W}(x,p;t)>0$) are shown.
\label{fig:wigner_third}}
\end{figure}

A similar, but even richer, situation can be seen at $T_{rev}/3$, where there
are three distinct `mini-packets' in position space, and corresponding
interesting interference structure in momentum space, as shown in
Fig.~\ref{fig:xp_third}. The explicit form of the wavefunction near this
fractional revival time is given by \cite{averbukh}
\begin{equation}
\psi(x,t\approx T_{rev}/3)
=
-\frac{i}{\sqrt{3}}
\left[ \psi_{cl}(x,t)
+
e^{2\pi i/3}\left\{
\psi_{cl}(x,t+T_{cl}/3)
+
\psi_{cl}(x,t+2T_{cl}/3)
\right\}\right]
\end{equation}
and exactly this type of highly correlated state is obvious in the
Wigner function visualization in Fig.~\ref{fig:wigner_third}, with
three smooth `lumps' corresponding to the diagonal terms in
Eqn.~(\ref{two_gaussian_wigner}), and three oscillatory
`cross-terms' at the average value locations in phase space
predicted by Eqn.~(\ref{average_values}), with the appropriate
`wiggliness'.

\begin{figure}
\epsfig{file=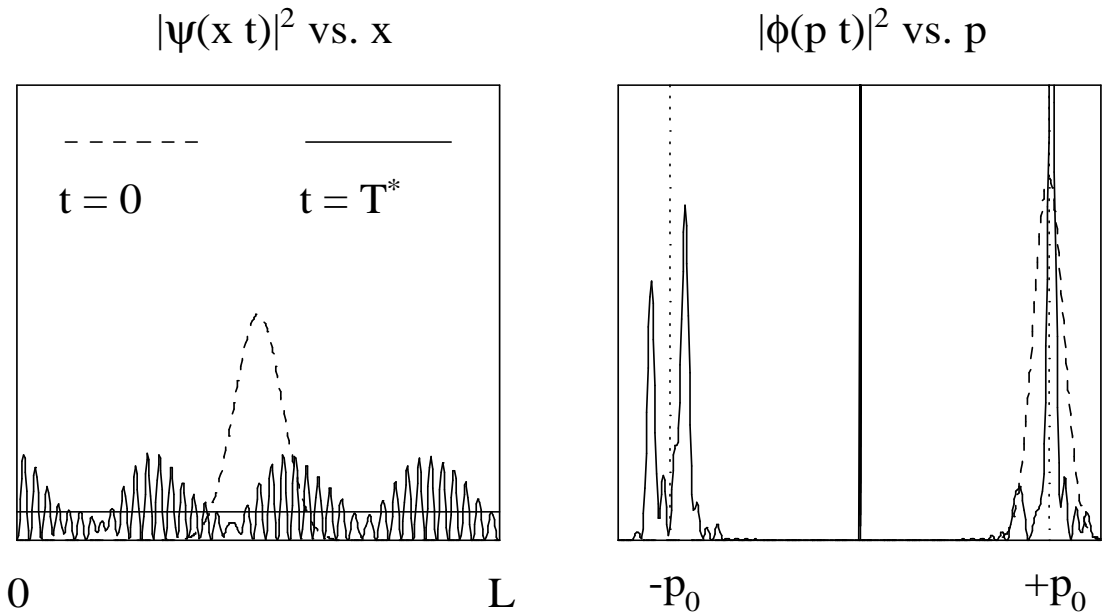,width=8cm,angle=270}
\caption{Same as Fig.~\ref{fig:xp_early}, except for $t=0$ (dashed)
and for a general time, $T^{*} = 16T_{rev}/37$ (solid), during the collapsed 
phase, not near any resolvable fractional revival.
\label{fig:xp_collapse}}
\end{figure}

\begin{figure}
\epsfig{file=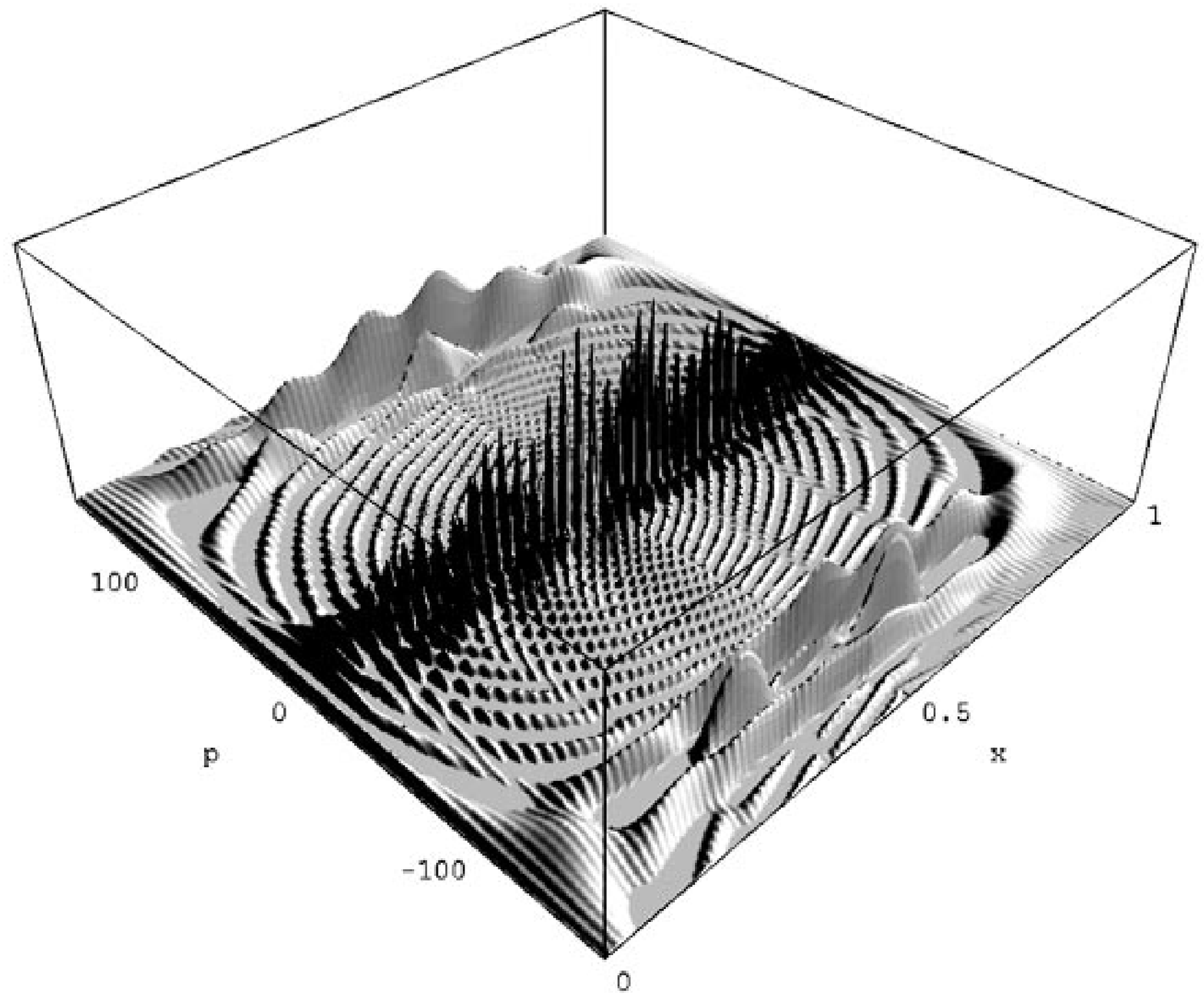,width=18cm,angle=0}
\caption{Same as Fig.~\ref{fig:wigner_early},  but for a general time
($T^{*} = 16T_{rev}/37$) during the collapsed phase, not near any resolvable 
fractional revival, to be compared to Fig.~\ref{fig:xp_collapse}.
Only positive values ($P_{W}(x,p;t)>0$) are shown.
\label{fig:wigner_collapse}}
\end{figure}

Finally, at other `random' times during the time evolution of such states, 
not near any resolvable fractional revivals, the wavefunction collapses to 
something like an incoherent sum of eigenstates, with little or no obvious 
correlations. A view of this behavior at such a time (represented here by 
$t = T^{*} = 16T_{rev}/37$) using $|\psi(x,t)|^2$ and $|\phi(p,t)|^2$ is 
shown in Fig.~\ref{fig:xp_collapse},  as well as using the Wigner function 
in Fig.~\ref{fig:wigner_collapse}.

\section{Conclusions and discussions}
\label{sec:conclusions_and_discussions}

While the infinite well is one of the most standard model problems in
all of introductory quantum mechanics, it continues to be used as a
benchmark system for the analysis and visualization of new quantum
phenomena, including wave packet revivals. In a similar vein, the
Wigner function, which was invented more than 70 years ago, continues to
play an important role in the analysis, understanding, and visualization
of quantum systems, and in  related fields such as signal analysis and
quantum optics. We
have provided a thorough review of both the analytic structure of the Wigner
function for energy eigenstates as well as for time-dependent Gaussian wave
packet states for this important exemplary quantum system, focusing on the
visualization of wave packet states at fractional revival times.
(Additional images not included in this paper can be accessed at
{\tt http://webphysics.davidson.edu/mjb/wigner/}.)

Given the relative straightforwardness of the calculations involved
here for the ISW, it is easy to imagine extending the results presented
here to simple extensions, such as a finite square well, or an asymmetric
infinite well of the form
\begin{equation}
      V(x) = \left\{ \begin{array}{ll}
+\infty & \mbox{for $x<-b$} \\
0  &  \mbox{for $-b<x<0$}  \\
+V_0  &  \mbox{for $0<x<+a$}  \\
+\infty  &  \mbox{for $+a<x$}
\end{array} \right.
\, .
\end{equation}
This last form is interesting as the eigenstates  exhibit less trivial
correlations between the magnitude/wiggliness of the position-space
wavefunction (due to the varying classical speeds in the two sides of the
well) which can then be connected to the behavior in momentum space more
directly using the Wigner function. This example is also instructive since,
because of the unphysical discontinuity of the potential, there are typically
surprising results \cite{robinett_asymmetric} when one compares quantum
results to classical expectations for position- and momentum-space
probability densities. This type of asymmetric well is also of the form
for which coherent charge oscillations have been observed experimentally
\cite{coherent_charge}, consisting of a two state system,
\begin{equation}
\psi(x,t) = \gamma \left[u_{n}(x) e^{-iE_at/\hbar} \right]
          + \delta \left[u_{b}(x) e^{-iE_bt/\hbar}\right]
\end{equation}
with $E_a<V_0$ and $E_b > V_0$.

\vskip 1cm

\noindent {\bf Acknowledgments} We would like to thank Wolfgang
Christian and Tim Gfroerer for useful conversations regarding this
work. MB was supported in part by a Research Corporation Cottrell
College Science Award (CC5470) and the National Science Foundation
(DUE-0126439).

\appendix
\section{Trigonometric integrals}
\label{appendix:trig_integrals}

For many of the calculations included here, we require versions of
the integral
\begin{equation}
\int_{-\infty}^{+\infty} \frac{\sin(z)\,\cos(mz)}{z}\,dz
=
\left\{ \begin{array}{ll}
0     & \mbox{for $m<-1$ and $m>+1$} \\
\pi/2 & \mbox{for $m=\pm 1$} \\
\pi   & \mbox{for $m^2<1$} \\
\end{array} \right.
\label{general_integral}
\end{equation}
which is a handbook \cite{math_handbook} result. We can briefly justify
these results, starting with the single integral
\begin{equation}
\int_{-\infty}^{+\infty} \frac{\sin(z)}{z}\,dz =  \pi
\end{equation}
which itself can be derived using contour integration.
This can be generalized to
\begin{equation}
\int_{-\infty}^{+\infty} \frac{\sin(mz)}{z}\,dz
=
\left\{ \begin{array}{ll}
+\pi     & \mbox{for $m>0$} \\
0 & \mbox{for $m=0$} \\
-\pi   & \mbox{for $m<0$} \\
\end{array} \right.
\label{simpler_integral}
\end{equation}
by a change of variables, and considering the special $m=0$ case
separately. The general integral in Eqn.~(\ref{general_integral})
can then  be obtained by writing
\begin{equation}
\sin(z) \cos(mz) =
\frac{1}{2} \left\{
\sin[(1+m)z]  + \sin[(1-m)z]
\right\}
\end{equation}
and using Eqn.~(\ref{simpler_integral}) twice. The special case of
$m=1$ is done by noting that
\begin{equation}
\int_{-\infty}^{+\infty}
\frac{\sin(z)\,\cos(z)}{z}\,dz
=
\int_{-\infty}^{+\infty} \frac{\sin(2z)}{2z}\,dx
=
\frac{1}{2}
\int_{-\infty}^{+\infty}
\frac{\sin(w)}{w}\,dw
= \frac{\pi}{2}
\end{equation}
where $w = 2z$ and a simple trig identity is used.
Another integral that  is useful for normalization calculations is
\begin{equation}
\int_{-\infty}^{+\infty}
\frac{\sin^2(z)}{z^2}\,dz = \pi
\end{equation}
which is another handbook result derivable using contour integration
techniques.

\section{Wigner distribution from momentum-space wavefunctions}
\label{appendix:wigner_momentum}

The evaluation of the Wigner function, using momentum-space wavefunctions as
in Eqn.~(\ref{wigner_definition_p}), requires the integral
\begin{eqnarray}
P_{W}^{(n)}(x,p) & = &
\left(\frac{L}{(\pi\hbar)^2}\right)
\int_{-\infty}^{+\infty}
\,dq\, e^{-2iqx/\hbar}
\,
e^{+i(p+q)L/2\hbar}
\,
e^{-i(p-q)L/2\hbar} \\
& &
\times
\left\{
e^{-in\pi/2}
\frac{\sin[((p+q)L/\hbar -n\pi)/2]}{[(p+q)L/\hbar -n\pi]}
-
e^{+in\pi/2}
\frac{\sin[((p+q)L/\hbar +n\pi)/2]}{[(p+q)L/\hbar +n\pi]}
\right\} \nonumber \\
& &
\times
\left\{
e^{+in\pi/2}
\frac{\sin[((p-q)L/\hbar -n\pi)/2]}{[(p-q)L/\hbar -n\pi]}
-
e^{-in\pi/2}
\frac{\sin[((p-q)L/\hbar +n\pi)/2]}{[(p-q)L/\hbar +n\pi]}
\right\}
\nonumber
\end{eqnarray}
and we briefly sketch out some of the necessary steps in the evaluation
of $P_{W}^{(n)}(x,p)$ in this approach.

Combining the various complex exponentials and using the fact that the
Wigner function must be real, we are left with integrals such as
\begin{equation}
{\cal I}_1 =
\int_{-\infty}^{+\infty}
\cos\left[\frac{qL}{\hbar}\left(\frac{L-2x}{L}\right)\right]
\,
\frac{\sin[((p+q)L/\hbar -n\pi)/2]}{[(p+q)L/\hbar -n\pi]}
\,
\frac{\sin[((p-q)L/\hbar -n\pi)/2]}{[(p-q)L/\hbar -n\pi]}
\, dq
\, .
\end{equation}
In this case, the appropriate partial fraction identity to
rewrite the denominators is
\begin{eqnarray}
 \frac{1}{((p+q)L/\hbar - n\pi)((p-q)L/\hbar - n\pi)}
& = &
\frac{1}{2(pL/\hbar -n\pi)}
\left\{
\frac{1}{((p+q)L/\hbar - n\pi)}
\right. \\
& &
\qquad  \qquad \qquad
\left.
+
\frac{1}{((p-q)L/\hbar - n\pi)}
\right\}
\nonumber
\end{eqnarray}
and the remaining integrals can be done using variations on
Eqn.~(\ref{important_integral}). Upon combining various factors, one
obtains integrals of that form not only with $m = (L-2x)/L$,  as
in Eqn.~(\ref{intermediate_integral}), but similar ones
with $m = (L-4x)/L$ and $m=(3L-4x)/L$. These terms give rise to
limits on the $x$ dependence of the form ${\cal R}(x;0,L/2)$ and
${\cal R}(x;L/2,L)$ where ${\cal R}(x;a,b)$ is defined in
Eqn.~(\ref{restricted_x_function}). For example, one intermediate
result can be written in the form
\begin{eqnarray}
{\cal T} & = &
\cos\left[\left(\frac{pL}{\hbar} - n\pi\right)\left(\frac{L-2x}{L}\right)
\right] \sin \left[\left(\frac{pL}{\hbar}-n\pi \right)\right]
{\cal R}(x;0,L) \\
& &
-
\sin\left[\left(\frac{pL}{\hbar} - n\pi\right)\left(\frac{L-2x}{L}\right)
\right] \cos \left[\left(\frac{pL}{\hbar}-n\pi \right)\right]
\left\{ {\cal R}(x;0,L/2) - {\cal R}(x;L/2,L) \right\}
\nonumber
\end{eqnarray}
which, in turn, gives
\begin{equation}
{\cal T} = \left\{ \begin{array}{ll}
0 & \mbox{for $x<0,x>L$} \\
\sin[(2(p/\hbar -n\pi/L)x]  & \mbox{for $0<x<L$} \\
\sin[(2(p/\hbar -n\pi/L)(L-x)]  &  \mbox{for $L/2<x<L$}
\end{array} \right.
\end{equation}
so that the `split' definition of $P_{W}^{(n)}(x,p)$ in the two half
intervals arises very naturally and the complete result in
Eqn.~(\ref{isw_wigner_distribution}) is reproduced using momentum-space
methods, including the non-trivial $x$ dependence.

\section{Problems}

\noindent
{\bf P1}: Complete the proof that the $\phi_{n}(p)$ for the ISW
are orthonormal by explicit calculation of
$\langle \phi_{m}|\phi_{n}\rangle$, completing the steps in
Sec.~\ref{sec:infinite_well_results}, and making use of identities
such as Eqn.~(\ref{useful_identity}).

\vskip 0.5cm

\noindent
{\bf P2}: Using results from any quantum mechanics textbook, evaluate
the Wigner distribution for the ground state and first excited state
of the simple harmonic oscillator. {\bf Answer}: The harmonic oscillator
eigenstates can be written in the form
\begin{equation}
u_{n}(z) = A_n H_n(z) e^{-z^2/2}
\quad
\mbox{where}
\quad
A_n \equiv \frac{1}{\sqrt{2^n n! \sqrt{\pi}}}
\quad
\mbox{and}
\quad
z \equiv \frac{x}{b} = \frac{x}{\sqrt{\hbar/m\omega}}
\end{equation}
and the $H_n(z)$ are the Hermite polynomials of order $n$. The Wigner
functions for $n=0,1$ are then given by
\begin{eqnarray}
P_{W}^{(0)}(x,p) & = & \frac{1}{\pi \hbar} e^{-\rho^2} \\
P_{W}^{(1)}(x,p) & = & \frac{1}{\pi \hbar} e^{-\rho^2} (2\rho^2-1)
\end{eqnarray}
where
\begin{equation}
\rho^2 \equiv \frac{x^2}{b^2} + \frac{b^2 p^2}{\hbar^2}
\,.
\end{equation}
This is one of the simpler results which shows explicitly that
the Wigner function need not be positive-definite. Note that
the first of these results is consistent with the expression for the
free-particle Gaussian in Eqn.~(\ref{free_particle_gaussian_wigner})
for vanishing $x_0$, $p_0$, and $t$. The Wigner function for arbitrary $n$ has
been  evaluated
(see, {\it e.g.}, Ref.~\cite{wigner_review}) with the result that
\begin{equation}
P_{W}^{(n)} = \frac{(-1)^n}{\pi \hbar} e^{-\rho^2}
L_{n}(2\rho^2)
\end{equation}
where $L_n(z)$ are the Laguerre polynomials.

\vskip 0.5cm

\noindent
{\bf P3}: Show how the phase-space plot for the 1-D harmonic oscillator
can be used to generate the classical probability densities for position-
and momentum-variables, namely, show how one obtains
\begin{equation}
P_{CL}(x) = \frac{1}{\pi \sqrt{x_A^2 -x^2}}
\qquad
\qquad
\mbox{and}
\qquad
\qquad
P_{CL}(p) = \frac{1}{\pi \sqrt{p_A^2 -p^2}}
\, .
\label{sho_final_answer}
\end{equation}
{\bf Partial answer}: The classical probability distribution from
Eqn.~(\ref{classical_sho_distribution}) can be written in the form
\begin{equation}
P_{CL}(x,p)
\quad
\propto
\quad
\delta(p(x)^2 - p^2)
= \frac{1}{2p(x)}\left[\delta(p(x)-p) + \delta(p(x)+p)\right]
\label{partial_sho_result}
\end{equation}
where
\begin{equation}
p(x) = \sqrt{2mE - m^2\omega^2 x^2} = m\omega \sqrt{x_A^2 - x^2}
\, \, .
\end{equation}
Integration of Eqn.~(\ref{partial_sho_result}) over $dp$ then gives
\begin{equation}
P_{CL}(x) \propto \frac{1}{p(x)} \propto \frac{1}{\sqrt{x_A^2 -x^2}}
\end{equation}
which when properly normalized (integrated over the classically allowed
interval $(-x_A,+x_A)$) gives the first term in Eqn.~(\ref{sho_final_answer}).

\vskip 0.5cm

\noindent
{\bf P4}: Using either the momentum- or position-space wavefunctions
in Eqns.~(\ref{p_gaussian_t}) or (\ref{x_gaussian_t}) for the
free-particle Gaussian
wave packet, show that the Wigner function is of the form in
Eqn.~(\ref{free_particle_gaussian_wigner}).

\vskip 0.5cm

\noindent
{\bf P5}: Complete the proof that the result for the
Wigner function for eigenstates of the ISW in
Eqn.~(\ref{isw_wigner_distribution})
can be obtained using momentum-space wavefunctions, using the techniques
in Appendix~\ref{appendix:wigner_momentum}.

\vskip 0.5cm

\noindent
{\bf P6}: Use the results in Eqn.~(\ref{time_dependent_wigner}) and
(\ref{off_diagonal_wigner}) to write down the time-dependent Wigner
function for the simple two-state system in the infinite well
\begin{equation}
\psi(x,0) = \frac{1}{\sqrt{2}}\left[u_{1}(x) + u_{2}(x)\right]
\end{equation}
and generate plots of $P_{W}^{(\psi)}(x,p;t)$ for various times.
Compare the results to standard images of the position-space and
momentum-space probability densities for this problem
\cite{robinett_book}, \cite{dean}. What is the only time scale associated
with this system? and does it have anything to do with a classical
periodicity?

 \newpage

\clearpage

\clearpage

\clearpage

\clearpage

\clearpage

\clearpage

\clearpage

\clearpage

\clearpage

\clearpage

\clearpage

\clearpage

 \clearpage

\clearpage

 \clearpage

%%%%%%%%%%%%%% we use 16*TREV/37 for the time %%%%%%%%%%%%%%%%%%%

 \clearpage

\begin{thebibliography}{99}
\bibitem{drawback}
A positive definite quasi-probability distribution can be obtained
by the use of a Gaussian `smearing' function in the definitions in
Eqns.~(\ref{wigner_definition_x}) or (\ref{wigner_definition_p}),
which gives rise to the so-called Husimi function \cite{wigner_review},
\cite{lee}, \cite{ballentine}, $P_{H}(x,p;t)$.
This quasi-probability density, however, does not give the correct marginal
distributions as in Eqns.~(\ref{integrate_p}) and (\ref{integrate_x}).
\end{thebibliography}
\end{document}